\documentclass[aps,prb,noshowpacs, superscriptaddress,twocolumn,float-fix,10pt,final,times]{revtex4}
\usepackage[pdftex]{hyperref}
\usepackage{amssymb} 
\usepackage{graphicx}
\usepackage{bm}
\usepackage{amsmath}
\usepackage{wrapfig}
\usepackage{float}
\usepackage{color}
\usepackage{times}
\usepackage{textgreek}
\usepackage{subfiles}

\def\ba{\begin{eqnarray}}
\def\ea{\end{eqnarray}}
\def\beq{\begin{equation}}
\def\eeq{\end{equation}}
\usepackage{subfiles}

\begin{document}

\title{Emergence of stationary many-body entanglement\\ in driven-dissipative Rydberg lattice gases}
\author{Sun Kyung Lee}
\address{Spin Convergence Research Center, Korea Institute of Science and Technology, Seoul 136-791, Korea}
\author{Jaeyoon Cho}
\address{School of Computational Science, Korea Institute for Advanced Study, Seoul 130-722, Korea}
\author{K. S. Choi}
\address{Spin Convergence Research Center, Korea Institute of Science and Technology, Seoul 136-791, Korea}
\address{Institute for Quantum Computing and Department of Physics \& Astronomy, University of Waterloo, Waterloo, Ontario N2L 3G1, Canada}

\begin{abstract}
{ {Non-equilibrium quantum dynamics represents an emerging paradigm for condensed matter physics, quantum information science, and statistical mechanics. Strongly interacting Rydberg atoms offer an attractive platform to study driven-dissipative dynamics of quantum spin models with long-range order. Here, we explore the conditions under which stationary many-body entanglement persists with near-unit fidelity and high scalability. In our approach, coherent many-body dynamics is driven by Rydberg-mediated laser transitions, while atoms at the lattice boundary reduce the entropy of the many-body state. Surprisingly, the many-body entanglement is established by continuously evolving a locally dissipative Rydberg system towards the steady-state, as with optical pumping. We characterize the dynamics of multipartite entanglement in a 1D lattice by way of quantum uncertainty relations, and demonstrate the long-range behavior of the stationary entanglement with finite-size scaling, reaching ``hectapartite" entanglement under experimental conditions. Our work opens a route towards dissipative preparation of many-body entanglement with unprecedented scaling behavior.}}
\end{abstract}

\maketitle 

\noindent Quantum control of \textit{open} many-body systems has become a major theme in the quest to explore new physics at the interface between condensed matter physics, quantum information science, and statistical mechanics \cite{Amico2008, Diehl2008, Verstraete2009,Campisi2011, Kastoryano2013}. The ability to control the many-body interactions and their dissipative processes has been identified as a powerful resource for the preparation of steady-state entanglement \cite{Plenio1999,Schneider2002,Plenio2002,Bruan2002, Jakobczyk2002, Munschik2011, Kastoryano2011, Cho2011} and the investigation of noise-driven quantum phase transitions \cite{Diehl2008}. Indeed, quantum reservoir engineering provides the framework for dissipative quantum computation \cite{Verstraete2009, Kastoryano2013} and communication \cite{Vollbrecht2011} with built-in fault-tolerance. Furthermore, open system dynamics offers new prospectives to the relationship between entanglement and quantum thermodynamics \cite{Campisi2011}. 

Laser-driven Rydberg atoms offer unique possibilities for creating and manipulating open quantum systems $\hat{\rho}$  of dipolar interacting spin models \cite{Jaksch2000, Lukin2001, Saffman2010RMP}. By exciting atoms to high-lying Rydberg states, strong and long-range interactions between the Rydberg atoms can be exploited to induce spin-spin interactions, whereas atoms comprising the many-body state can couple to their local radiative reservoirs by spontaneous emission \cite{Rydbergbook}. The competition between the coherent and incoherent dynamics can drive the system to bipartite entangled states for two atoms \cite{Rao2013,Carr2013} and novel states of matter for a mesoscopic number of atoms, exhibiting topological order, glassiness, and crystallization dynamics \cite{Weimer2010,Lee2012, Zhao2012, Ates2012,Glaetzie2012,Honing2013,Lesanovsky2013}. Remarkably, the basic primitives behind such a principle have been demonstrated in the laboratory by several groups \cite{Wilk2010,lsenhower2010, Schau2012,Dudin2012, Peyronel2012,Schempp2013}. 

Despite the tantalizing prospects of quantum-reservoir engineering, the main obstacle has been that local decoherence (e.g., spontaneous emission) generally destroys the global entanglement of the system. Most proposals reported to date thereby achieve the required ``non-local" jump operator by way of collective system-bath coupling \cite{Plenio1999,Schneider2002,Plenio2002,Bruan2002, Jakobczyk2002, Munschik2011, Kastoryano2011, Cho2011} in order to suppress the information loss by local dissipation. In practice, such a coupling is achieved in the highly challenging, strong coupling regime for an array of qubits interacting with a common reservoir (e.g., cavity mode). Furthermore, the inherently local nature of the driving fields hardly allows only a single entangled state to be distinctively separated from the coupling to the reservoir, which enforces the introduction of auxiliary coherent manipulations and multiple time-steps of quantum gates and dissipations to single out a particular entangled state from a broader subspace\cite{Weimer2010}, diluting the very nature of quantum-reservoir engineering. 

Another challenge is the characterization of entanglement in the many-body state $\hat{\rho}(t)$  under evolution \cite{Amico2008,Guhne2009}. For interacting spin systems, spin waves are the quasiparticle excitations describing the ``collective"  state of the atomic mode. Entanglement in such a system can be defined by the correlations among the collective excitations \cite{Amico2008}. Hence, verification protocols for mode entanglement can be extended to extract the many-body entanglement of these quantum spin systems \cite{Sorensen2001, Duan2011, Logouvski2009, Papp2009}. Uncertainty relations, as defined in Refs. \citen{Logouvski2009, Papp2009}, can be applied to access the \textit{genuine} multipartite entanglement of $\hat{\rho}$ (Ref. \citen{Choi2010}).

\begin{figure}[tH!]
\includegraphics[width=0.9\columnwidth]{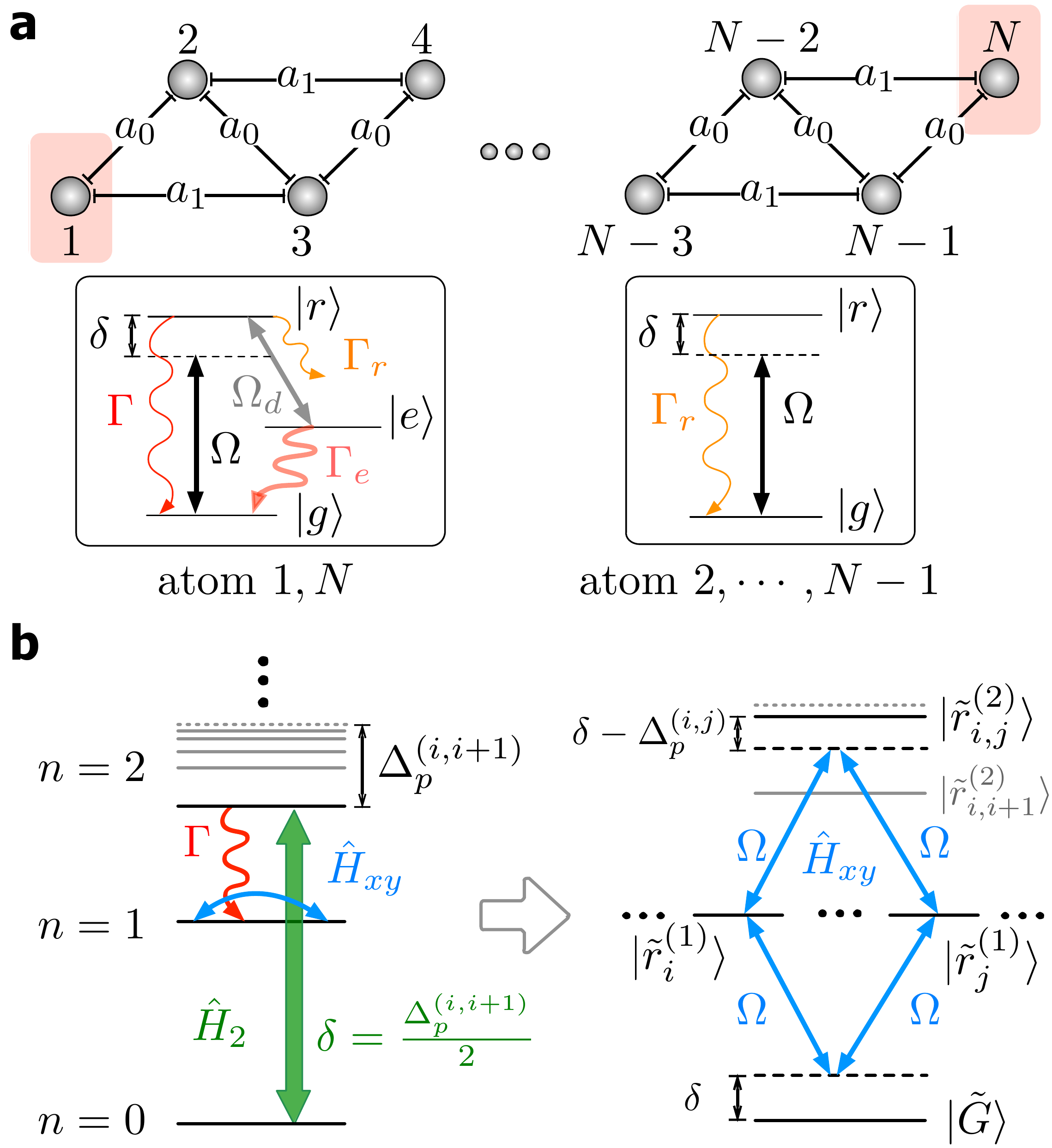}\\ 
\caption{\textbf{Production of stationary entanglement with Rydberg atoms in 1D lattice.} \textbf{(a)} Schematic of optically-driven, dissipative Rydberg atoms in a 1D staggered triangular lattice. Distances $a_0$ and $a_1$ are defined between spins in neighbor and next-neighbor configurations. Inset \textbf{(i)} The decay rates  $\Gamma_{1,N}=\Gamma\gg\Gamma_r$  for the edge atoms are enhanced by mixing the Rydberg states $|r\rangle$ with short-lived $|e\rangle$ with fields $\Omega_d$ (Methods). Inset \textbf{(ii)} Atoms are pumped by a driving field $\Omega$ with detuning $\delta$. \textbf{(b)} Rydberg blockaded atomic structure showing a rich family of anharmonic levels separated by subspace $n$. (Left) Two-photon process $\hat{H}_2$ optically pumps the population to the $n=1$ manifold. (Right)  XY Hamiltonian $\hat{H}_{xy}$ dictates the delocalization dynamics within the $n=1$ subspace. $J_{ij}$ is driven by Raman transitions $|\tilde{r}_i^{(1)}\rangle\leftrightarrow |\tilde{r}_j^{(1)}\rangle$, while $\bar{\Delta}_\text{ls}^{(i)}$ is induced by light shifts (Methods).
}\label{fig1}
\end{figure}

Here, we explore such many-body entangled states persisting with high fidelity in the stationary limit for laser-driven Rydberg atoms in a lattice. As illustrated in Fig. \ref{fig1}, our protocol conceptually begins by globally pumping regularly-arranged Rydberg atoms $(A\oplus B)$ with a driving field $\Omega$, where the lattice is separated into two partitions $A$, $B$. Rydberg excitation coherently delocalizes within the subspace defined by ``system" atoms $A$, while ``reservoir" atoms $B$ at the lattice boundary serve as an entropy sink for $A$ with local fields $\Omega_d$ that enhance the spontaneous decay. By preparing a dark state in the Markovian dynamics, the atomic sample evolves towards the entangled steady-state 
in the form of an eigenstate $|\epsilon_1\rangle=|W\rangle_A \otimes |g\cdots g\rangle_B$ of a many-body Hamiltonian $\hat{H}_{xy}$ in the single-excitation subspace, where $|W\rangle_A$ $(|g\cdots g\rangle_B)$ is a $W$-like entangled state (ground state) for $A$ ($B$). 

We apply our method to generate bipartite entanglement for $N=4$ atoms, and extend our work for $N=6$ to investigate the driven-dissipative dynamics of the many-body entanglement with the uncertainty relations \cite{Logouvski2009, Papp2009}. We find that quadripartite $W$-state persists indefinitely with fidelity $F\geq 0.99$ for $N=6$, and that entanglement depth $k$ shows favorable scaling relative to its system size, reaching ``hectapartite" ($k=100$) entanglement for $N=128$ atoms. Unlike previous methods with auxiliary unitary and time-sequential manipulations \cite{Plenio1999,Schneider2002,Plenio2002,Bruan2002, Jakobczyk2002, Munschik2011, Kastoryano2011, Cho2011,Weimer2010}, the many-body entanglement emerges purely out of the open system dynamics in a time-independent, continuous fashion with local decoherence, as with the original spirit of optical pumping. Our method thereby allows the scalable production of high-fidelity dissipative many-body entanglement with Rydberg atoms. 

\vspace{0.25cm}

{\noindent\fontfamily{phv}\selectfont \textbf{Results}}

\noindent\textbf{Schematics of optically-driven, dissipative Rydberg atoms.} We consider the many-body states of $N$ atoms configured in a lattice [See  Fig. \ref{fig1}(a)], irradiated by a uniform driving field  $\Omega$  that couples the atomic ground state $|g\rangle$ to the highly excited Rydberg state $|r\rangle$ with detuning $\delta$. A pair of atoms $i,j$ in the Rydberg state at lattice sites $\vec{x}_i,\vec{x}_j$ couple each other via the potential $\Delta^{(ij)}_p=C_p |\vec{x}_i-\vec{x}_j|^{-p}$ with power-law scaling, for which we take $p=6$ for the van der Waals regime of blockade shifts \cite{Rydbergbook}. In a frame rotating with the laser frequency, the Hamiltonian is given by 
\begin{equation}
\hat{H}= \sum_{i=1}^{N}\left (\delta\hat{\sigma}^{(i)}_{rr} +\Omega\hat{\sigma}_x^{(i)}\right)-\sum_{\langle i, j \rangle}^{N}\Delta^{(ij)}_p\hat{\sigma}_{rr}^{(i)} \hat{\sigma}_{rr}^{(j)},
\label{eq1}
\end{equation} 
where $\hat{\sigma}_{\mu\mu}^{(i)}=|\mu\rangle_i \langle \mu |$ is the projection operator for states $|\mu\rangle$ with $\mu\in\{g,r\}$, and $\hat{\sigma}_k^{(i)}$ are the canonical Pauli operators for atom $i$ with $k\in\{x,y,z,\pm\}$. $\langle i,j\rangle$ denotes the sum over all $i\neq j$. In the following, we denote the ground state ($n=0$) as $|\tilde{G}\rangle=|g\cdots g\rangle$, the singly-excited ($n=1$)  states as $|\tilde{r}_i^{(1)}\rangle=|g_1\cdots r_i \cdots g_N\rangle$, and the doubly-excited ($n=2$)  states as $|\tilde{r}_{ij}^{(2)}\rangle=|g_1\cdots r_i\cdots r_j\cdots g_N\rangle$ for the excitation subspace $n=\sum_i\langle\hat{\sigma}_{rr}^{(i)}\rangle$.

The open many-body  dynamics for the atomic state $\hat{\rho}$ is governed by a Markovian master equation $\dot{\hat{\rho}}=-i[\hat{H},\hat{\rho}]+{\mathcal{L}}\hat{\rho}$ with the Lindblad superoperators $\mathcal{L}\hat{\rho}=\sum_{i}\frac{\Gamma_{i}}{2}(\hat{\sigma}^{(i)}_{+}\hat{\rho}\hat{\sigma}^{(i)}_{-}-\{\hat{\sigma}^{(i)}_{+}\hat{\sigma}^{(i)}_{-},\hat{\rho}\})$ for the atomic coupling to their local radiative reservoirs. As discussed in the Methods, in order to allow the jump $n\rightarrow n-1$, we can arbitrarily set the decay rate $\Gamma_i\simeq |\Omega_d|^2/\Gamma_e$ relative to its free-space rate $\Gamma_r$ by coherently mixing the Rydberg level $|r\rangle$ and a rapidly decaying $|e\rangle$ with field $\Omega_d$, where $\Gamma_e$ is the decay rate of $|e\rangle$ (inset of Fig. \ref{fig1}).  

\vspace{0.25cm}

\noindent\textbf{Dissipative production of many-body entanglement.} As shown by Fig. \ref{fig1}(b), our dissipative protocol starts by optically pumping the population into $n=1$ subspace by driving $|\tilde{G}\rangle\rightarrow |\tilde{r}_{i,i+1}^{(2)}\rangle$ with the field $\Omega$ through two-photon transition $\hat{H}_2$ with $\delta=\Delta_p^{(i,i+1)}/2$. Higher-order transitions ($n=1\rightarrow n=3$) are suppressed for moderate $N$ due to the long-range nature of $\Delta_p^{(ij)}$ (Methods). We thereby adiabatically eliminate $|\tilde{G}\rangle$ and $|\tilde{r}_{ij}^{(2)}\rangle$ in the off-resonant limit $|\delta-\Delta_p^{(ij)}|\gg w_d$, and obtain an effective Hamiltonian
\begin{equation}\label{Hxy}
\hat{H}_{{xy}}=\sum_{\langle i,j\rangle}  J_{ij} \left ( \hat{\sigma}^{(i)}_{+} \hat{\sigma}^{(j)}_{-}+\hat{\sigma}^{(i)}_{-} \hat{\sigma}^{(j)}_{+} \right)-\sum_{i=1} \bar{\Delta}_{\text{ls}}^{(i)}\hat{\sigma}_{rr}^{(i)}
\end{equation}
in the single-excitation manifold, where $w_d=\sqrt{\Gamma_r^2/4+2\Omega^2}$ is the power-broadened linewidth. The XY Hamiltonian $\hat{H}_{{xy}}$ delocalizes the Rydberg excitation between sites $i,j$ at a ``hopping" rate $J_{ij}=\tfrac{\Omega^2}{\delta}-\tfrac{\Omega^2}{\delta-\Delta_p^{(ij)}}$, with each site $i$ subjected to a ``magnetic" field $\bar{\Delta}_{\text{ls}}^{(i)}=\tfrac{\Omega^2}{\delta}-\sum_{j\neq i}\tfrac{\Omega^2}{\delta-\Delta_p^{(ij)}}$ (Methods). 

\begin{figure}[t!]
\begin{center}
\includegraphics[width=0.8\columnwidth]{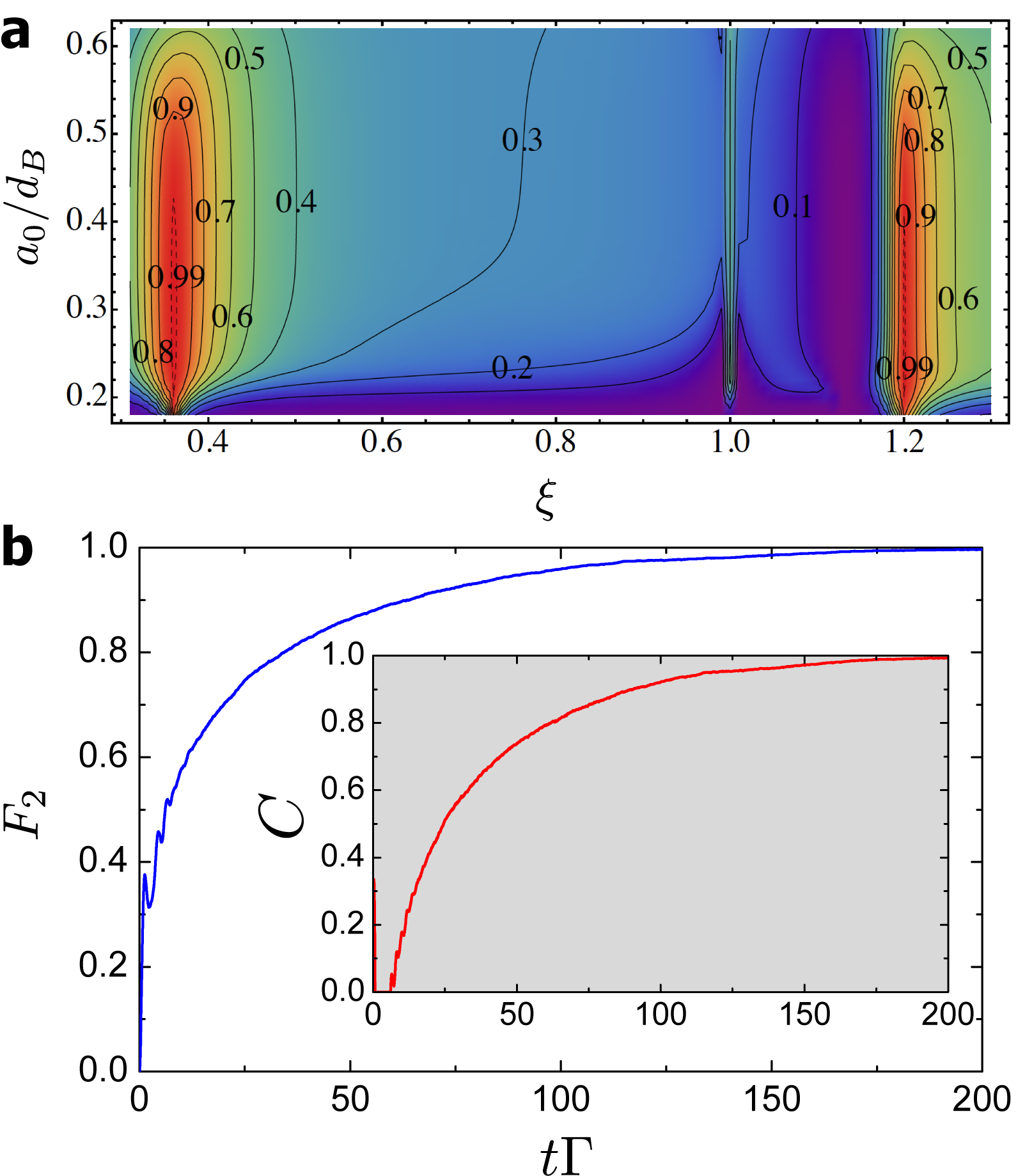}
\caption{\textbf{Driven-dissipative dynamics of bipartite atomic entanglement.} \textbf{(a)} Contour of stationary entanglement fidelity $F_2$ with interaction parameter $\xi$ and distance $a_0$ (in units of blockade radius $d_B$). \textbf{(b)} Dynamics of entanglement fidelity $F_2(t)$ as a function of pumping time (in units of $\Gamma$). Inset. Temporal evolution of concurrence $C$ from unentangled $C=0$ to maximal entanglement $C=1$ for the parameters: $\delta=w_d/2(a_0/d_B)^6$, $\Omega=10^3 \Gamma$, and $\Gamma_{1,4}=10^{4}\Gamma_r$ for atoms $1,4$ with $\{\xi,a_0/d_B\}=\{\sqrt[6]{3},0.26\}$. }
\label{fig2}
\end{center}
\end{figure}

The dissipative many-body entanglement for the steady-state $\lim_{t\rightarrow\infty}\hat{\rho}=\hat{\rho}_{ss}$ is prepared as follows. We first identify the spectrum $\{\epsilon_i,|\epsilon_i\rangle\}$ of $\hat{H}_{xy}$ in the $n=1$ subspace. Our goal is to set $J_{ij},\bar{\Delta}_{\text{ls}}^{(i)}$ such that one and only one of the eigenstates, say $|\epsilon_1\rangle$, corresponds to a product of $W$ state $|W\rangle_A=\sum_{i\in A}|\tilde{r}_i^{(1)}\rangle$ for a subset $A$ of atoms (``system atoms") and ground state $|g\cdots g\rangle_B$ for another subset $B$ (``reservoir atoms"), thereby leading to $|\epsilon_1\rangle=|W\rangle_A \otimes |g\cdots g\rangle_B$. We control the relative hopping rates $J_{i,i\pm x}$ between nearest neighbors ($x=1$) and next-nearest neighbors ($x=2$) in a lattice to obtain dark resonance for atoms $B$. By enhancing $\Gamma_i$ for atoms $B$, the atomic sample is dissipatively driven to the entangled dark state $|\epsilon_1\rangle$. 

\vspace{0.25cm}

\noindent\textbf{Emergence of dark states for open many-body dynamics.} Qualitatively, the dark resonances $J_{i,i+1}= -J_{i,i+2} $ occur for atoms $\{1,N\}$ at the lattice boundary of a 1D staggered triangular lattice in Fig. \ref{fig1}(a) for $\xi= \sqrt[6]{3}$ and $N=4$. $\xi$ determines the relative strength between nearest ($a_0$) and next-nearest ($a_1$) neighbor interactions by the relation $\xi=a_1/a_0$. More generally, for $N\gg 4$, quantum interference between multiple pathways $|\tilde{r}_i^{(1)}\rangle\leftrightarrow|\tilde{r}_j^{(1)}\rangle$ occurs so that $|\epsilon_1\rangle=|W\rangle_A \otimes |g\cdots g\rangle_B$ emerges as the unique dark state  (Methods). This process is analogous to coherent population trapping (CPT) for levels consisting of ``radiative" states $\{ |\tilde{r}^{(1)}_{i\in B}\rangle\}$ with decay rate $\Gamma_{i\in B}\gg\Gamma_r$ coupled to ``metastable" states $\{ |\tilde{r}^{(1)}_{i\in A}\rangle\}$. We define atoms $B$ as suitable atomic reservoirs, whereby the atoms are continuously projected to the ground state by spontaneous emission $\Gamma$ (Ref. \citen{Lewenstein1995}). In order to enable this process, we locally enhance the decoherence $\Gamma\simeq |\Omega_d|^2/\Gamma_e$ for the reservoir atoms $B$ by $\simeq 10^4$ relative to the radiative rates $\Gamma_r$ of the system atoms $A$.  Any Rydberg population in atoms $B$ will cause the overall atomic state to become ``bright" and decay until it reaches the unique steady-state $|\epsilon_1\rangle$. Many-body entanglement is thereby established for the stationary state $\hat{\rho}_{ss}=|\epsilon_1\rangle\langle \epsilon_1 |$.

The entanglement dynamics displays an intricate behavior, as the atomic sample is driven to the steady-state $\hat{\rho}_{ss}$. At the early stage of Liouvillian dynamics ($0\leq t_1\leq 1/\Gamma$), atoms in $|\tilde{G}\rangle=|g\cdots g\rangle$ are rapidly pumped to the $n=1$ subspace. The Rydberg excitation then delocalizes under $\hat{H}_{xy}$ with off-resonant Raman transitions $J_{ij}$. At the final stage ($t_2\gg 1/\Gamma$), the Rydberg lattice gas $\hat{\rho}$ is dissipatively pumped to a $W$-like entangled state $|W\rangle_A$, which separates from $|g\cdots g \rangle_B$. The entanglement fidelity $F$ is thereby determined by the ``branching" ratio $\Gamma_r/\Gamma\simeq 10^{-4}$ between the lifetimes of dissipative and coherent dynamics. Because our procedure does not involve adiabatic evolutions, our dark-state pumping protocol is in principle scalable to arbitrarily large $N$ with extended samples $L\gg d_B$ only limited by $F=1-\mathcal{O}(\Gamma_r/\Gamma)$. By continuously driving the system towards $\hat{\rho}_{ss}$, the many-body entanglement is auto-stabilized in the presence of noise and decoherence. 

\vspace{0.25cm}

\noindent\textbf{Open-system dynamics for bipartite atomic entanglement.} In the following, we perform a numerical analysis of the relaxation behavior of the Rydberg gas to a stationary bipartite entanglement for atom number $N=4$ and enhanced radiative rates $\Gamma_{1,N}=\Gamma$ for the edge atoms. Fig. \ref{fig2}(a) displays the contour map of entanglement fidelity $F_2=\langle \psi_{2}| \text{Tr}_B[\hat{\rho}_{\text{ss}}] |\psi_{2}\rangle$ for the stationary state $ \hat{\rho}_{\text{ss}} $ relative to $|\psi_{2}\rangle=1/\sqrt{2}(|g_2 r_3\rangle+|r_2 g_3\rangle)$ as a function of interaction parameter $\xi$ and distance $a_0$ (in units of blockade radius $d_B=\sqrt[6]{C_6/w_d}$). The profile of fidelity along $a_0$ depicts the requirement of Rydberg-blockade regime $a_0<d_B$ to provide sufficient nonlinearity in $n$ [Fig. \ref{fig1}(b)], selectively driving transitions $|\tilde{G}\rangle\leftrightarrow|\tilde{r}_{i,i+1}^{(2)}\rangle$ and adiabatically eliminating subspaces $n=0,2$ (Methods). Atoms in the region $0.2\leq a_0/d_B\leq 0.5$ are thereby efficiently pumped to the single-excitation subspace. The interaction parameter $\xi$ is tuned to numerically maximize the steady-state entanglement fidelity up to $F_2=0.9982$ for $\xi_1=\sqrt[6]{3}$ and $\xi_2=0.36$ at $a_0/d_B=0.26$. To validate our entanglement pumping scheme, we further show the dissipative dynamics of concurrence $C$ (Ref. \citen{Guhne2009}) for $\xi_1$ in the inset of Fig. \ref{fig2}(b). The atomic sample is driven to a maximally entangled state with $F_2=0.9965$ within $t\Gamma=200$.

\vspace{0.25cm}

\begin{figure}[t!]
\begin{center}
\includegraphics[width=0.9\columnwidth]{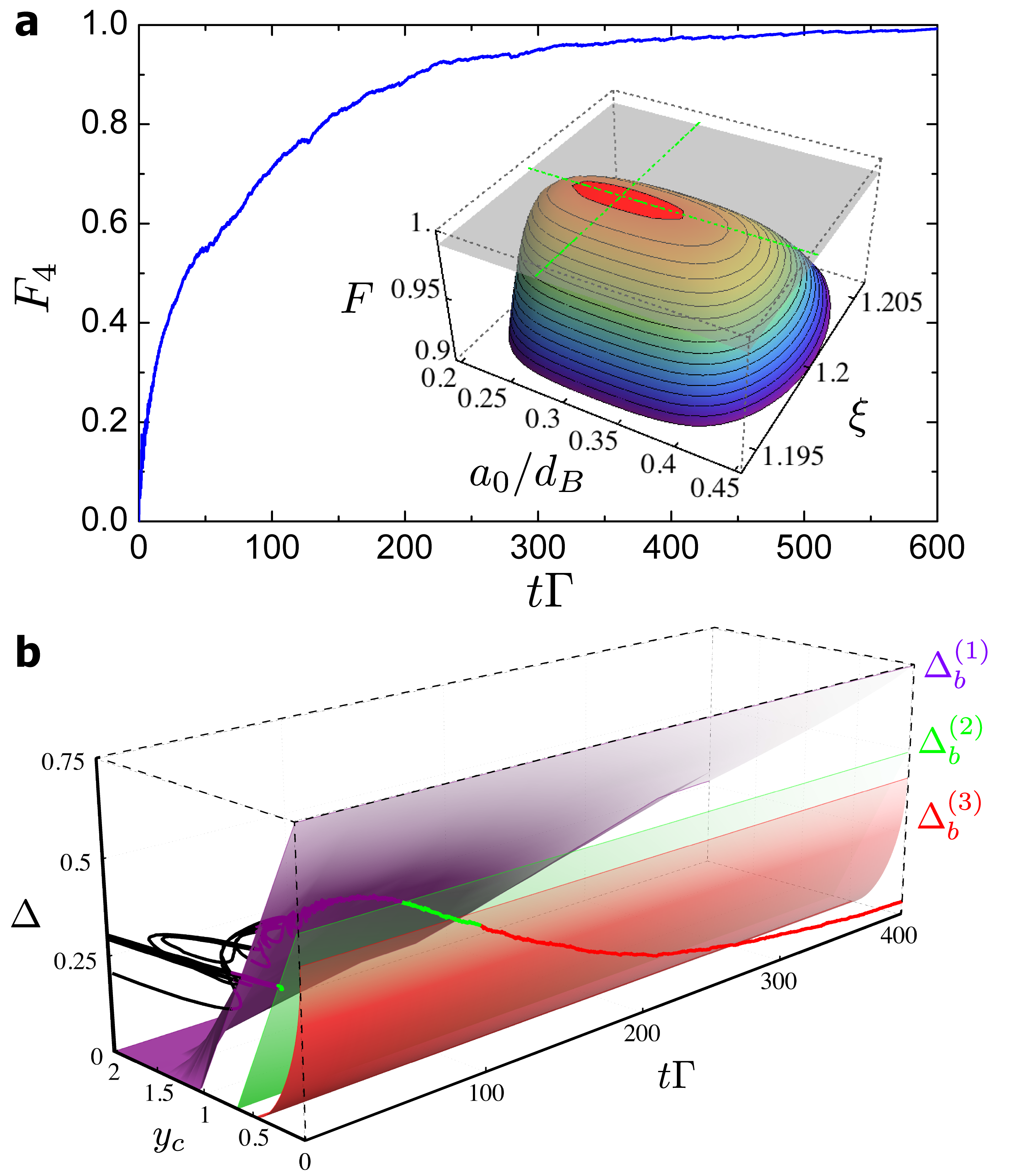}
\caption{\textbf{Driven-dissipative dynamics of many-body entanglement for six atoms.} \textbf{(a)} Dynamics of entanglement fidelity $F_4$ at maximum point $\{\xi, a_0/d_B\}=\{1.1996, 0.285\}$ simulated by way of Monte-Carlo wavefunction. Inset. 3D map of steady-state entanglement fidelity $F_4$ for interaction parameter $\xi$ and distance  $a_0$ (in units of blockade radius $d_B$). $F_4>0.99$ for $0.25\leq a_0\leq 0.35$ \textbf{(b)} Dissipative preparation of genuine quadripartite entangled state. The entanglement parameters $\{\Delta(\rho), y_c\}$ transit from fully separable  (black) to bipartite entanglement  (purple, $t_2\Gamma\simeq 2$), to tripartite entanglement (green, $t_3\Gamma\simeq 65$), and to stationary quadripartite entanglement (red, $t_4\Gamma\simeq 100$).  }\label{fig3} 
\end{center}
\end{figure}

\noindent\textbf{Evolution of many-body entanglement and uncertainty-based entanglement witness.} Now, let us treat the case of many-body entanglement with $N=6$ atoms in the 1D lattice, as an example of multipartite system. With same parameter set $\Omega$ and $\Gamma$, we simulate the dissipative dynamics of entanglement fidelity $F_4(t)=\langle \psi_{4}| \text{Tr}_B [\hat{\rho}(t)] |\psi_4\rangle$, with respect to the ideal symmetric $W$ state $|\psi_4\rangle=\frac{1}{2}\sum_{i=2}^5 |\tilde{r}_i^{(1)}\rangle$ by way of quantum-trajectory method [See Fig. \ref{fig3}(a)]. Here, we have optimized the steady-state fidelity $\max(F_4)=0.9912$ for the parameters $\{\xi,a_0/d_B\}=\{1.1996,0.285\}$, thereby setting a symmetric quadripartite $W$-state $|\epsilon_1\rangle=|\psi_4\rangle\otimes |g_1, g_6\rangle$.

The dissipative transitions of genuine many-body entanglement is detected by the uncertainty relations \cite{Logouvski2009, Papp2009,Choi2010}, which serves as the collective entanglement witness $\{\Delta(t),y_c(t)\}$ (Ref. \citen{Amico2008}). The uncertainty $\Delta=\sum_i \langle\delta^2\hat{\Pi}_i\rangle$ measures the total variance of projection operators $\hat{\Pi}_i=|W_i\rangle\langle W_i |$ to $N_A$-dimensional $W$-state basis $|W_i\rangle$, while $y_c=\tfrac{2N_A}{N_A-1} \tfrac{p_{\geq 2} p_{0}}{p_1^2}$ detects the amount of higher-order spin-waves (e.g., $p_2=\sum_{i\neq j} \langle\hat{\sigma}_{rr} ^{(i)}\hat{\sigma}_{rr} ^{(j)}\rangle$) and ground-state fraction $p_0=\sum_i\langle\hat{\sigma}_{gg} ^{(i)}\rangle$ relative to the singly-excited spin wave $p_1=\sum_{i} \langle\hat{\sigma}_{rr} ^{(i)}\rangle$, where $N_A$ is the number of atoms in $A$. For an ideal $W$-state, $\min\{{\Delta},{y}_c\}\rightarrow\{0,0\}$, while the boundary $\Delta_b^{(k-1)}$ represents the minimum uncertainty for $(k-1)$-partite entangled states for a given $y_c$. Violation of the uncertainty bound $\Delta(\hat{\rho})<\Delta_b^{(k-1)}$ then signals the presence of genuine $k$-partite entanglement stored in $\hat{\rho}(t)$, with the full $N_A$-partite entanglement certified by $0\leq \Delta(\hat{\rho})<\Delta_b^{(N_A-1)}$. $\{\Delta,y_c\}$ can be measured by the transverse collective spin variance and by the excitation statistics (Methods).

By applying the witness $\{\Delta,y_c\}$, we observe that atoms initially in ground state are dissipatively driven to the quadripartite entangled $W$ state by sequentially crossing the boundaries $\Delta_b^{(1)}, \Delta_b^{(2)},\Delta_b^{(3)}$ in Fig. \ref{fig3}(b). The dissipative transitions of many-body entanglement are indicated by black, purple, green, and red lines of Fig. \ref{fig3}(b) for the average trajectory  $\hat{\rho}(t)$. For pumping time $t_4 \sim 100/\Gamma$, the many-body system exhibits a full quadripartite entanglement with a moderate atom number $N=6$, and reach $\{\Delta,y_c\}|_{ss}\rightarrow\{1.5\times 10^{-2},2\times10^{-4}\}$, as the atoms are pumped to the desired eigenstate $|\epsilon_1\rangle$.

\begin{figure}[t]
\begin{center}
\includegraphics[width=1\columnwidth]{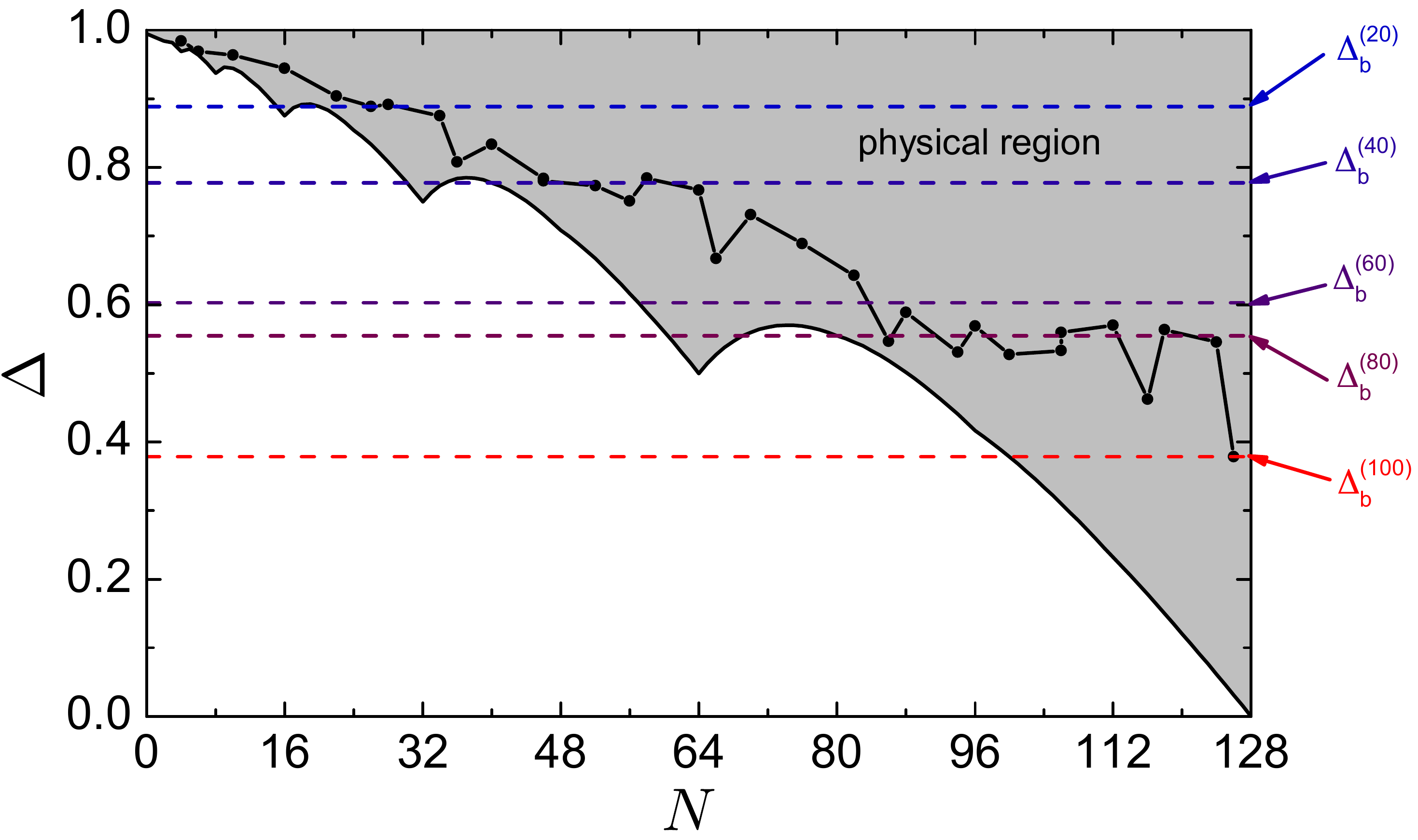}
\caption{\textbf{Finite-size scaling behavior of many-body entanglement depth.} Multipartite entanglement behavior of the many-body system $\hat{\rho}_{ss}$ is probed with quantum uncertainty witness $\Delta$ for $y_c\rightarrow 0$ by way of direct diagonalization of $\hat{H}_{xy}$ as a function of atom number $N$. We obtain stationary eigenstates  $\hat{\rho}_{ss}=|\epsilon_1\rangle\langle \epsilon_1 |$, exhibiting up to hectapartite entanglement for $N=128$ atoms. The uncertainty boundaries for $20$-partite, $40$-partite, $60$-partite, $80$-partite, $100$-partite entanglement are shown as dashed lines.}
\label{fig4} \end{center}
\end{figure}

\vspace{0.25cm}

\noindent\textbf{Finite-size scaling of steady-state entanglement.} 
Next, we move on to the question of finite-size scaling behavior of the stationary many-body entanglement. Although the full dynamical simulation for large $N$ is beyond our computational capability, the steady-state entanglement can be established by analyzing the unique eigenstate $|\epsilon_1\rangle$ that meets the dark resonance condition $\xi=\xi_1$, for which $J_{i,i+1}=-J_{i,i+2}$. Perturbations by higher-order interactions are negligible, as $\sum_{x> 2}^{\infty} |J_{i,i+x}|/|J_{i,i+1}|\ll 10^{-2}$. We truncate our analysis up to next-nearest-neighbor interactions for the following discussion. We define the entanglement depth $k$ in accord with the concept of $k$-producibility for qubits \cite{Amico2008,Guhne2009}, thereby identifying the minimal depth for genuine $k_m$-partite entanglement to produce the purported state $\hat{\rho}_{ss}$. 

We directly diagonalize the many-body Hamiltonian $\hat{H}_{xy}$ for $\xi_1$, and characterize the resulting entanglement depth $k$ of the stationary eigenstate $|\epsilon_1\rangle$ up to $N\rightarrow 128$. Fig. \ref{fig4} captures our result of $\{\Delta,y_c\rightarrow 0\}$ for the dark state $|\epsilon_1\rangle=|W_k\rangle_A  \otimes |g\cdots g\rangle_B$, where $|W_k\rangle_A$ is the $k$-partite symmetric $W$ state.  Due to the nonlinear sensitivity of our witness for some region $k$, we characterize the scaling of the \textit{minimal} entanglement depth $k_m\leq k$ (Methods). The shaded area represents the physical region, whereby $k_m$-partite entanglement could be defined for a given $N$, and the dashed lines are the uncertainty bounds for $0\sim 100$-partite entanglement (with $20$-partite increments). Remarkably, we observe a favorable scaling up to genuine ``hectapartite" ($k_m=100$) entanglement for $N=128$ atoms.

\vspace{0.25cm}
{\noindent\fontfamily{phv}\selectfont \textbf{Discussion}}

\noindent Our entanglement pumping scheme is experimentally feasible. By exciting $^{85}$Rb atoms to Rydberg state $|r\rangle=|100S_{1/2}\rangle$, quadripartite entangled states could be prepared for $F_4>0.99$ within $t_4=10$ $\mu s$ in the region $1(1.2)\mu\text{m}\leq a_0 (a_1)\leq 1.5 (1.8) \mu\text{m}$. The limit for any driven-dissipative approach with Rydberg lattice gases will be the photoionization lifetime $t_{\text{ph}}\gg 4$ ms (Methods). Since the pumping time to reach $\hat{\rho}_{ss}$ depends on $N$, our method can be applied to generate stationary hectapartite entanglement within $t_p<t_{\text{ph}}$ for $N=128$ atoms with $a_0\simeq 1\mu$m  (Methods). 

For $N\gg 128$, one could explore emergent atom-field systems embedded in photonic crystals. Dispersive optical interactions near band edges can induce dipole-dipole oscillations $\hat{H}_{xy}$ and ``Rydberg" blockades $\hat{H}_{2}$ with tailored scaling $\Delta_p^{(ij)}\sim c^{-x_{ij}}$ between low-lying excited atoms \cite{Shahmoom2013, Douglas2013}. Decay rates $\Gamma_{i}=\Gamma^{\prime}$ can be controlled by the density of states \cite{Goban2013}. More generally, the delocalization dynamics $\hat{H}_{xy}$ in the high-order subspace $n$ (Methods) can be extended to examine locality estimates of many-body systems \cite{Lieb-Robinson, Eisert2013} and bosonic sampling for quantum algorithms \cite{Childs2013}.  

We have examined the conditions under which driven-dissipative dynamics displays a rich family of many-body entangled states, and have provided a criteria for the purported entanglement. The stationary many-body entanglement shows a favorable long-range behavior up to $k_m=100$ for $N=128$ atoms. Our work thereby opens the door towards an open system simulator with well-controlled coherent and dissipative many-body dynamics, monitored by information-based quantities \cite{Amico2008,Verstraete2009,Diehl2008}.

This work is funded by the KIST Institutional Programs, and, in part, by the Ontario Ministry of Research \& Innovation and Industry Canada. We acknowledge the support of NVIDIA Corporation with equipment donations.

\vspace{1cm}
{\noindent\fontfamily{phv}\selectfont \textbf{Methods}}
\vspace{0.25cm}
\small{

\noindent\textbf{Control of spontaneous emission rates.}\label{contspontsection}
As discussed in the main text, for reservoir sites $i\in B$, atoms initially in the Rydberg state $|r\rangle$ with decay rate $\Gamma_r$ radiatively couple to a highly decoherring state $|e\rangle$ with decay rate  $\Gamma_e\gg\Gamma_r$ so that atoms in bipartition $B$ can behave as an effective ``reservoir" channel for the ``system" atoms in partition $A$.  In this section, we discuss how we could manipulate the spontaneous emission rate $\Gamma_{i}$ of the Rydberg state $|r\rangle$ for the ``reservoir" atoms. 

As illustrated in Supplementary Fig. 1(a), we consider a $\Lambda$-type energy level diagram, where $|r\rangle$ is dressed with $|e\rangle$ by auxiliary field $\Omega_d$. In the rotating-wave frame of the dressing laser $\Omega_d$, the Hamiltonian is  given by 
\begin{equation}
\hat{H}_d=\Delta_d \hat{\sigma}_{ee}^{(i)}+\Omega_d\left (\hat{\sigma}_{er}^{(i)}+\hat{\sigma}_{re}^{(i)}\right).
\end{equation}
The resulting optical Bloch equations are, then, 
\begin{eqnarray}
\dot{{\sigma}}_{ge}^{(i)}&=&-\gamma_e{\sigma}_{ge}^{(i)}+i\Delta_d\sigma_{ge}^{(i)}+i\Omega_d{\sigma}_{gr}^{(i)}\label{HL1}\\
\dot{{\sigma}}_{gr}^{(i)}&=&-\gamma_r{\sigma}_{gr}^{(i)}+i\Omega_d{\sigma}_{ge}^{(i)},\label{HL2}
\end{eqnarray}
where $\gamma_{e,r}=\Gamma_{e,r}/2$ and $\Delta_d$ is the detuning for the dressing field $\Omega_d$ relative to the transition $|e\rangle\leftrightarrow |r\rangle$. In writing Eqs. \ref{HL1}--\ref{HL2}, we have neglected the Langevin noise forces $\hat{F}_{\mu\nu}$  and  assumed $c$-number counterparts for $\hat{\sigma}_{\mu\nu}^{(i)}\mapsto {\sigma}_{\mu\nu}^{(i)}$. Hence, we find that the atomic coherence ${{\sigma}}_{gr}^{(i)}(t)$ between  $|g\rangle, |r\rangle$ obeys the following equation of motion
\begin{equation}
\ddot{\tilde{\sigma}}_{gr}^{(i)}-(i\tilde{\Delta}+\gamma_r)\dot{\tilde{\sigma}}_{gr}^{(i)}e^{(-i\tilde{\Delta}-\gamma_r)t}+\Omega_d^2 \tilde{\sigma}_{gr}^{(i)}e^{(-i\tilde{\Delta}-\gamma_r)t}=0,
\end{equation} 
with $\tilde{\sigma}_{gr}^{(i)}=\sigma_{gr}^{(i)}e^{-\gamma_r t}$ and $\tilde{\Delta}=i\gamma_e+\Delta_d$.

Supplementary Fig. 1(b) shows the dynamics of Rydberg population ${\sigma}_{rr}^{(i)}(t)$ obtained by numerically solving Eqs. \ref{HL1}--\ref{HL2} for the parameters of Figs. 2--4 with $\Gamma_e=10^4\Gamma_r$. The black solid (dashed) line is the atomic dynamics for $\Omega_d=10\Gamma_r$  ($\Omega_d\in\{10^2\Gamma_r,\cdots,9\times 10^2\Gamma_r\}$ with $10^2\Gamma_r$ increments). The red line is the result of atomic decay $\Gamma\simeq 10^3\Gamma_r$ with $\Omega_d=10^3\Gamma_r$. As we increase $\Omega_d\rightarrow \Gamma_e $, we find that the effective decay rate for the reservoir atoms scales with $\Gamma\sim |\Omega_d|^2/\Gamma_e$ up to $\Omega_d\sim 0.1 \Gamma_e$. 

In order to understand the dynamics, we formally integrate Eq. \ref{HL1} to obtain $\sigma_{ge}^{(i)}e^{-i\tilde{\Delta}t}=i\Omega_d \int \sigma_{gr}^{(i)}e^{-i\tilde{\Delta} t} dt \simeq \frac{\Omega_d}{\tilde{\Delta}}\sigma_{gr}^{(i)} e^{-i\tilde{\Delta} t}$. Assuming slowly-varying amplitude $\dot{\sigma}_{gr}^{(i)}$ for $\Omega_d\ll \gamma_e$, we obtain the following equation of motion
\begin{equation}
\dot{\sigma}_{gr}^{(i)}=-\left (\gamma_r+\frac{i\Omega_d^2}{\tilde{\Delta}} \right ) \sigma_{gr}^{(i)},
\end{equation}
where the effective decay rate is given by $\gamma_{\text{eff}}=\gamma_r+\frac{\gamma_e |\Omega_d|^2 }{|\Delta|^2+\gamma_e^2}$ with $\Gamma=2\gamma_{\text{eff}}$. As further discussed below, $|r\rangle=|100S_{1/2}\rangle$ and $|e\rangle=|5P_{1/2}\rangle$ have decay rates with $\Gamma_e/\Gamma_r\simeq 10^4$. Hence, decay rates for reservoir sites could be enhanced up to $4$ order of magnitude with $\Gamma/\Gamma_r\rightarrow 10^4$.

\vspace{0.25cm}

\noindent\textbf{Optical pumping to arbitrary $n$-subspace in an anharmonic Rydberg ladder.} Now, let us discuss the possibility of optically pumping the system $\hat{\rho}$ of $N$ atoms to an arbitrary target $n_t$-excitation subspace with $n_t<N-2$, for which $n_t=1$ in the main text. This is achieved by a set of $n_t$ lasers resonantly driving the two-photon transitions $n\rightarrow n+2$ ($n \in \{0,\cdots, n_t-1\}$) with effective Rabi frequencies $\Omega_2^{(n)}$ [See Supplementary Fig. 2(a)] and the three-photon transition  $n_t-2\rightarrow n_t+1$ with effective Rabi frequency $\Omega_3^{(n)}$ [see Supplementary Fig. 2(b)]. Because  $\mathcal{L}\hat{\rho}$ dissipates the levels $n\rightarrow n-1$, the atomic population is pumped to the target subspace $n_t$ [See Supplementary Fig. 2(b)]. For the case of $n_t=1$, $\Omega_2^{(0)}$ is provided by a single global field $\Omega$ for the entire atoms [See Supplementary Fig. 2(c)]. 

The efficacy of this procedure to address only a particular transition $n\rightarrow n^{\prime}$ depends on the anharmonicity in the Rydberg spectrum $V_n=\langle n |\hat{V}_p | n \rangle$, where $\hat{V}_p=\sum_{\langle i,j\rangle}^{N}\Delta_{p}^{(ij)}\hat\sigma_{rr}^{(i)}\hat\sigma_{rr}^{(j)}$ and $|n\rangle$ represents the most shifted state of the $n$ subspace. The $V_n$ is obtained by degenerate Rydberg configurations with $n$-nearest neighbor excitations (e.g., $|n\rangle=| r_1,\cdots,r_n,g_{n+1},\cdots, g_N\rangle$). The Rydberg spectrum is then given by
\begin{equation}
V_n=\sum_{i=1}^{n-1}\sum_{j=i+1}^{n}\Delta_p^{(ij)}.
\end{equation}
The transition energy for $n\rightarrow n+2$ is then
\begin{equation}
V_{n+2}-V_{n}=2\sum_{i=1}^{n}\Delta_{p}^{(i,n+1)}+\Delta_{p}^{(1,n+2)},
\end{equation}
so that the anharmonicity is given by
\begin{equation}
\delta V_{n+2,n}=\Delta_{p}^{(1,n+1)} +\Delta_{p}^{(1,n+2)}.
\end{equation} 

As shown in Supplementary Fig. 2(b), for a given target subspace $n_t$, we terminate the two-photon excitations to $n_t-1\rightarrow n_t+1$. All subspaces with $n \in\{0,\cdots,n_t-1,n_t+1\}$ are resonantly connected by two-photon transitions $\Omega_2^{(n)}$ with detunings $\delta_n^{(2)}=(V_{n+2}-V_{n})/2$ and by three-photon $\Omega_3$ coupling with detuning $\delta_{n_t-2}^{(3)}=(V_{n_t+1}-V_{n_t-2})/3$, except for the $n_t$ subspace [See Supplementary Fig. 2 (b)]. The Rydberg blockade condition for the two-photon transition $n\rightarrow n+2$ is then given by 
\begin{equation}
\delta V_{n+2,n}> w_d^{(2)}, \label{genblockade}
\end{equation}
where $w_d^{(2)}=\sqrt{\Gamma_r^2+2|\Omega_2^{(n)}|^2}$ is the power-broadened width of the ``two-photon" transition $n\rightarrow n+2$ and $\Omega_2^{(n)}=2\Omega^2/\delta_n^{(2)}$ is the effective Rabi frequency. 

\vspace{0.25cm}

\noindent\textbf{Optical pumping to the single-excitation subspace.} For $n_t=1$, by driving the two-photon transition $n=0\rightarrow n=2$ with $\delta_0^{(2)}=(V_2-V_0)/2=\Delta_p^{(1,2)}/2$, the atoms are pumped to a decoherence-free subspace (DFS) for atoms $A$ of the $n_t=1$ subspace [see Supplementary Fig. 2(d)]. As discussed in the main text, the DFS is defined by the space spanned by superpositions of $\{|\tilde r_{i\in A}^{(1)} \rangle\}$, and the subspace (DS) is set for the reservoir atoms $B$. In this case, high pumping efficiency to $n_t=1$ is assured if the higher-order transition $n=1\rightarrow n=3$ is blockaded for the least shifted state $|r_1,r_2,g_3,\cdots,g_{N-1},r_N\rangle$ of $n=3$ subspace, thereby $\Delta_{p}^{(1,N-1)} +\Delta_{p}^{(2,N-1)}>w_d^{(2)}$. For the 1D lattice in Fig. \ref{fig1}(a), our dissipative pumping scheme works in the region $a_0/d_B\simeq 0.13$ even for $N\geq 100$, where we take $\Omega=10^2\Gamma$, $\Gamma/\Gamma_r=10^4$ and $\xi=a_1/a_0=\sqrt[6]{3}$. For $|r\rangle=|100S_{1/2}\rangle$, the blockade distance is $d_B=5.8\mu $m, so that $(a_0,a_1)\simeq (750\text{ nm},900\text{ nm})$.

\vspace{0.25cm}

\noindent\textbf{Derivation of effective spin Hamiltonian.}
In the off-resonant limit $|\delta-\Delta_p^{(ij)}|\gg w_d$, we obtain the effective Hamiltonian $\hat{H}_{\text{eff}}$ (Eq. \ref{Hxy}) by truncating the perturbative expansion to the second order and by time-averaging highly oscillating terms \cite{James2007},
\begin{eqnarray}
\hat{H}_{\text{eff}} &=& \sum_{m,n} \frac{[\hat{h}_m^{\dagger},\hat{h}_m] }{\bar{\omega}_{mn}} e^{i(\omega_m-\omega_n)t}\nonumber\\
&+& \sum_{m,n}\left (\frac{\hat{h}_n \hat{h}_m e^{-i(\omega_m+\omega_n)t}}{\bar{\omega^\prime}_{mn}}+h.c\right ),
\end{eqnarray}
with the interaction Hamiltonian given by
\begin{equation}
\hat{H}_{\text{I}}= \sum_{n=1}^{N} \hat{h}_n e^{-i\omega_n t}+\hat{h}_n^\dagger e^{i\omega_n t},
\end{equation}
where $H_I=e^{iH_0t}H_1e^{-iH_0t}$, $\bar\omega_{mn}=[({1}/{2})({1}/{\omega_m}+{1}/{\omega_n})]^{-1}$, and $\bar{\omega^\prime}_{mn}=[({1}/{2})({1}/{\omega_m}-{1}/{\omega_n})]^{-1}$. In particular, we use 
\begin{eqnarray}
\hat{H}&=&\hat{H}_0+\hat{H}_1\\
\hat{H}_0&=&\sum_{i=1}^{N} \delta \hat{\sigma}_{rr}^{(i)} -\sum_{i< j }^{N}\Delta_{p}^{(ij)}\hat\sigma_{rr}^{(i)}\hat\sigma_{rr}^{(j)}\nonumber\\
\hat{H}_1&=&\Omega\sum_{i=1}^{N}( \hat\sigma_{rg}^{(i)}+\hat\sigma_{gr}^{(i)}),\nonumber
\end{eqnarray} 
with $\hat\sigma_{\mu\nu}^{(i)}=|\mu\rangle_i\langle\nu |$ for $\mu,\nu \in \{g,r\}$ and blockade shift $\Delta_p^{(ij)}=C_p | \vec{x}_i-\vec{x}_j|^{-p}$.  

We obtain the following effective Hamiltonian
\begin{equation}
\hat{H}_{xy}=-\sum_{i=1} \bar{\Delta}_{\text{ls}}^{(i)}\hat{\sigma}_{rr}^{(i)} +\sum_{i<j}  J_{ij} \left ( \hat{\sigma}^{(i)}_{+} \hat{\sigma}^{(j)}_{-}+h.c\right ),
\end{equation}
which corresponds to a XY model $\hat{H}_{xy}$ with spin-spin interaction $J_{ij}$ and magnetic field $\Delta_{\text{ls}}^{(i)}$, thereby $\hat{H}_{\text{eff}}= \hat{H}_{xy}$. After the population is pumped to the $n_t$ subspace [see Supplementary Fig. 2(b)], the coherent atomic dynamics is governed by $H_{xy}$ within the $n_t$ subspace. 

For $n_t=1$, the necessary Raman couplings ($J_{ij}$) and light shifts ($\Delta_{\text{ls}}^{(i)}$) are generated by the global field $\Omega$ with detuning $\delta=\Delta_{p}^{(i,i+1)}/2$, for which
\begin{eqnarray}
\bar\Delta_{\text{ls}}^{(i)}&=&\frac{2\Omega^2}{\Delta_p^{(i,i+1)}}(1-\sum_{ j \neq  i }^{N}f_{ij}) \\
J_{ij} &=& \frac{2\Omega^2}{\Delta_p^{(i,i+1)}}(1-f_{ij}),
\label{eq:H_eff_c}
\end{eqnarray} 
where $f_{ij}= \left[1-\tfrac{\Delta^{(ij)}_p}{\Delta^{(i,i+1)}_p /2}\right]^{-1}$. The exchange term $J_{ij}$ involves Raman transitions between $|\tilde{r}_i^{(1)}\rangle$ and $|\tilde{r}_j^{(1)}\rangle$  through the ground states $|\tilde{G}\rangle$ with the $2\Omega^2/{\Delta_p^{(i,i+1)}}$ term, and through the $n=2$ manifolds $|\tilde{r}_{ij}^{(2)}\rangle$ with the $-2\Omega^2f_{ij}/{\Delta_p^{(i,i+1)}}$ term. The global field $\Omega$ also resonantly drives $n=0\rightarrow n=2$ transition with
\begin{equation}
\hat{H}_2=\sum_{i=1}  \frac{2\Omega^2}{\Delta_{p}^{(i,i+1)}} \left (\hat{\sigma}^{(i)}_{+} \hat{\sigma}^{(i+1)}_{+}+h.c \right).
\end{equation}
Since we have increased the decay rates $\Gamma_{i\in B}=\Gamma$, the population is driven to the $n_t=1$ subspace via $\hat{H}_2$ [see Supplementary Fig. 2]. As illustrated in the inset of Supplementary Fig. 3(a), the atomic dynamics in $n_t=1$ subspace is dictated by $H_{xy}$, whose coefficients are fully determined by the ratio $\Delta_{ij}/\Delta_{i,i+1}$ in a scale-invariant fashion (with overall factor $2\Omega^2/\Delta_{p}^{(i,i+1)}$). Generally, let us express the eigenstate $|\epsilon_{\mu}\rangle$ of $H_{xy}$ in $n_t=1$ as $|\epsilon_{\mu}\rangle=\sum_{i}\alpha_{i,\mu}|\tilde{r}_i^{(1)}\rangle$. 

\vspace{0.25cm}

\noindent\textbf{Emergence of dark states in the effective Hamiltonian.}
For certain values of $J_{ij}$, eigenstates $|\epsilon_{\mu}\rangle$ could include zero populated coefficients $\alpha_{i,\mu}=0$ for sites $i\in B$. We enhance the decay rates for these ``reservoir" atoms $B$ to set the dark resonance condition for $|\epsilon_{1}\rangle$. In the following, we show how this process could be mapped to a more familiar phenomena of coherent population trapping (CPT). This guiding principle allows to unambiguously set decoherring channels for the reservoir atoms $B$.  

Supplementary Fig. 2(d) depicts the manifold $n=1$ in the eigenbasis $\{ |\epsilon_\mu\rangle \}$ of $\hat{H}_{xy}$, instead of the usual $\{|r_{i}^{(1)}\rangle\}$. During the entanglement pumping, the population is constantly projected to some superposition state $|\epsilon^{\prime}(t)\rangle$ of eigenstates $\{ |\epsilon_\mu\rangle \}$ by the spontaneous decay channels $n=2\rightarrow n=1$ in atoms $B$.  If $|\langle\epsilon_1|\epsilon^{\prime}(t)\rangle|< 1$, the Rydberg population $|\tilde{r}_i^{(1)}\rangle$ will delocalize until it populates the reservoir atoms, thereby quickly decaying to $|\tilde G\rangle$ before being repumped by $\hat{H}_2$. After several cycles of $n=0\rightarrow n=2$ (via $\hat{H}_2$) and $n\rightarrow n-1$ (via $\Gamma$), the atomic population accumulates in the unique ``dark" eigenstate $|\epsilon_{1}\rangle$ of $\hat{H}_{xy}$. The steady-state entanglement for $|\epsilon_{1}\rangle $ is thereby continuously auto-stabilized by the balance of the rates $\Gamma, \Gamma_r$.
   
For the 1D staggered triangular lattice in Fig. \ref{fig1} (a), the position vectors are given by $\vec x_{i}=\{(k-1)a_1,0\}$ for odd sites ($i=2k+1$) and by $\vec x_{i}=\{a_0\cos\theta+(k-1)a_1,a_0\sin\theta \}$ for even sites ($i=2k$), with $\cos\theta=\xi/2$ and $\xi=a_1/a_0$. Under this geometry, the parameter $\xi$ can fully describe the effective Hamiltonian $H_{xy}$. Supplementary Fig. 3(b) shows the finite-range behavior of the nonlocal coupling rate $J_{ij}$ between $|\tilde r_{i}^{(1)}\rangle\leftrightarrow |\tilde r_{j}^{(1)}\rangle$ in the van der Waals (vdW) interacting regime ($p=6$). For the sufficiently large $\xi>1$, we find that the rate $J_{ij}$ significantly diminishes for sites $|i-j|>2$ due to the $\sim 1/r^{6}$ vdW scaling. In the following discussion, we thereby truncate our analysis up to next-nearest neighbor interactions with the sparse-array $H_{xy}$ as  
\begin{equation}
\hat{H}_{xy}=-\sum_{i=1}^{N} \bar{\Delta}_{\text{ls}}^{(i)}\hat{\sigma}_{rr}^{(i)} +\sum_{i<j}^{N}  J_{ij} \left ( \hat{\sigma}^{(i)}_{+} \hat{\sigma}^{(j)}_{-}+h.c\right ),
\end{equation}
with
\begin{eqnarray}
&&\bar\Delta_{\text{ls}}^{(i)} = \left\{
  \begin{array}{ll}
 \frac{J}{2}\times (4+\tfrac{2}{2-\xi^6}-N) & \text{for }i=1,N \\
 \frac{J}{2}\times(6+\tfrac{2}{2-\xi^6}-N) & \text{for }i=2,N-1 \\
 \frac{J}{2}\times(6+\tfrac{4}{2-\xi^6}-N) & \text{for }2<i<N-2\\
  \end{array}
  \right.\\
&&J_{ij} = \left\{
  \begin{array}{lr}
 J & \text{for }|i-j|=1 \\
J\times(\tfrac{1}{2-\xi^6}) & \text{for } |i-j|=2 \\
   0 & \text{for }|i-j|>2
  \end{array}
\right.
\end{eqnarray}
with overall factor $J=4\Omega^2/\Delta_{p}^{(i,i+1)}$.  

Eigenstates $|\epsilon_{\mu}\rangle$ with $\alpha_{i,\mu}=0$ can be obtained by controlling the ratio between nearest and next-nearest terms for $J_{ij}$ with $\xi$. As discussed in Supplementary Fig. 3 (a), this process is analogous to the behavior of CPT, where destructive quantum interference occurs for the excitation pathways that connects the ``bright" state $|\tilde{r}_{i}^{(1)}\rangle$ (decay rate $\Gamma\simeq 10^4 \Gamma_r$) to ``metastable" states $|\tilde{r}_{j\neq i}^{(1)}\rangle$  (decay rate $\Gamma_r$). The emergence of ``dark state" for such a toy model provides an insight on our choice of interaction parameter $\xi\rightarrow\xi_1= \sqrt[6]{3}$ for symmetric (antisymmetric) eigenstates, whereby $J_{i,i+1}=-J_{i,i+2}$. For instance, in the case of $N=4$ with
\begin{equation}
\hat{H}_{xy}=\left( \begin{array}{cccc}
-J & J &-J& 0\\
 J &0  & J &-J\\
 -J &J &  0  & J \\
0  & -J &J &-J\\
\end{array} \right),
\end{equation}
destructive interference in the form $J_{1,2}(J_{3,4})=-J_{1,3}(J_{2,3})$ occurs for $\alpha_{1}=\alpha_4=0$. For $N\gg 4$, the eigenstate $|\epsilon_1\rangle$ with $\alpha_{i\in B}=0$ cannot be obtained by locally considering the atoms near the boundaries (i.e., atoms $1,2,3$ and $N-2,N-1, N$). Instead, the uniqueness of the dark state $|\epsilon_1\rangle$ is a manifestation of the many-body interferences for $J_{ij}, \bar\Delta_{\text{ls}}^{(i)}$, leading to $\alpha_{i\in B}=0$. Nonetheless, $J_{i,i+1}=-J_{i,i+2}$ provides a reasonable guiding principle for us to guess the dark resonance conditions for atoms near the edges for a certain value of $N$, due to symmetric sparse characteristics of $H_{xy}$. 
 
We confirmed this prediction by solving the full spectrum of the sparse Hamiltonian matrix $\hat{H}_{xy}$ with $J_{|i-j|>2}\rightarrow 0$ and by numerically simulating the stationary state of the master equation. We obtain two sets of eigenstates $|\epsilon_{\mu}\rangle=\sum_{i} \alpha_{i,\mu} |\tilde{r}^{(1)}_{i}\rangle$ with $\alpha_{1,\mu}=\alpha_{N,\mu}=0$ for arbitrary $N$ that meets $J_{i,i+1}=-J_{i,i+2}$ at $\xi=\xi_1$ as below
\begin{eqnarray}
{\rm{set}}\,1&:&\, N=4+6m\,\,(m=0,1,\cdots)\label{set1eqs}\\
\{\alpha_{i,\mu}\}&=&\{ 0,1,1,0,-1,-1,0,1,1,0,\cdots,-1,-1,0,1,1,0\}  \nonumber \\
{\rm{set}}\,2&:&\, N=6+10m\,\,(m=0,1,\cdots)\label{set2eqs}\\
\{\alpha_{i,\mu}\}&=&\{ 0,1,1,1,1,0,-1,-1,-1,-1,0,1,1,1,1,0,\cdots \nonumber \\
&&-1,-1,-1,-1,0,1,1,1,1,0\}.  \nonumber 
\end{eqnarray} 
Therefore, our method could produce stationary $k$-partite entanglement in the form of an eigenstate $|\epsilon_{1}\rangle=|W_k\rangle_A\otimes |g\cdots g\rangle_B$ with $k=2+4m$ (set 1) and $k=4+8m$ (set 2), plotted as blue dots in Supplementary Fig. 4 (b). In the hindsight, we can attribute the existence of symmetric entangled steady-states in Eqs. \ref{set1eqs}--\ref{set2eqs} to the special structure of $\hat{H}_{xy}$. As $\hat{H}_{xy}$ is sparse, highly symmetric and redundant, a kind of commensurability requirement is imposed to the eigenstate under the restriction that the coefficients at the edges are zero.

Even if we were to consider for $^{\forall} J_{ |i-j|>2}$, our result would not have changed for $\xi=\xi_1$. The truncation would slightly modify the exact eigenstate as $|\epsilon_i\rangle\rightarrow |\epsilon_i\rangle+|\delta \epsilon_i\rangle$ with $\langle\delta\epsilon_i | \epsilon_i\rangle=0$ up to a normalization constant $\sim 1$. Roughly speaking, $\langle \delta\epsilon_i | \delta \epsilon_i\rangle$ scales linear to the perturbation in the energy scale $\sim 10^{-2}$ up to a leading order. Since the energy perturbation to $\Delta_p^{(ij)}\sim 1/r_{ij}^6$ is at most $\delta\Delta_p=\Delta_p^{(i,i+3)}/\Delta_p^{(i,i+1)}\simeq 10^{-2}$ [see Supplementary Fig. 3(b)], the perturbation to the entanglement fidelity is at most $\delta F\simeq 10^{-2}$, which is well within the numerical uncertainty of the quantum trajectory method [see Fig. \ref{fig3}]. In terms of dark resonance $J_{i,i+1}=-J_{i,i+2}$, the higher-order interactions $J_{|i-j|>2}$ for $\xi=\xi_1$ (blue dots) are suppressed by at least $10^2$ relative to $J_{i,i+1},J_{i,i+2}$. By taking $N\rightarrow \infty$, the higher-order contributions $\sum_{x>2}^{\infty} |J_{i,i+x}|$ would still be far too negligible to have any impact on the final state with $\sum_{x>2}^{\infty} |J_{i,i+x}|\ll 10^{-2}\times \min(|J_{i,i+1}|,|J_{i,i+2}|)$, leading to $F\simeq 0.99$. 

The infinitesimally reduced fidelity can then be recovered to $F\rightarrow 1$ by displacing $\xi$ to an optimal value by the more general condition $J_{1,2}=-\sum_{x\geq 2} \alpha_{x,1}J_{1,1+x}$ for $|\epsilon_1\rangle$ at the expense of having a slightly modified steady-states, i.e., a new eigenstate $|\epsilon_1^{\prime}\rangle=|W^{\prime}\rangle_A \otimes |g \cdots g \rangle_B$. Thanks to the inherent symmetry of the system, this modified steady-state $|W^{\prime}\rangle$ would only marginally differ from the original one. Furthermore, the original steady-state $|\epsilon_1\rangle$ could be recovered by re-adjusting the arrangement of the atoms. In any case, the only sensitive parameter that determines the optimal fidelity is the ``branching" ratio $\Gamma_r/\Gamma\simeq 10^{-4}$, which sets the balance between the lifetimes for dissipative and coherent evolutions, thereby the final fidelity $F\sim 1-\mathcal{O}(\Gamma_r/\Gamma)$.

On the other hand, in the region of $\xi\ll 1$ ($J_{|i-j|>2}\geq J_{|i-j|=1}, J_{|i-j|=2}$), the optimal value $\xi$ cannot be predicted by the dark resonance conditions of the sparse-array matrix $H_{xy}$. In this case, $J_{ij}$ displays zigzag oscillatory decay as shown in Supplementary Fig. 3(b), and higher-order terms $J_{|i-j|>2}$ must be included in the analysis. 

\vspace{0.25cm}

\noindent\textbf{N-partite uncertainty witness.} In this section, we describe our method of constructing the $N$-partite uncertainty witness from Ref. \citen{Logouvski2009}. Our entanglement witness $\{\Delta,y_c\}$ consists of identifying the boundaries $\Delta_b^{(k-1)}$ for all possible states $\hat{\rho}_b^{(k-1)}$ produced by convex combinations of pure $(k-1)$-partite entangled states $|\psi_b^{(k-1)}\rangle$ as well as their mixed siblings with less $k$. As shown in Refs. \citen{Hofmann2003,Logouvski2009}, the lower bound of $\Delta_b^{(k-1)}$ is attained by taking a convex set of $\{\Delta_b(\hat{\rho}_b^{(k-1)}),y_c(\hat{\rho}_b^{(k-1)})\}$ for all pure states $\hat{\rho}_b^{(k-1)}=|\psi_b^{(k-1)}\rangle\langle \psi_b^{(k-1)} |$. In Fig. \ref{fig3}, we depict the boundaries $\Delta_b^{(1)},\Delta_b^{(2)},\Delta_b^{(3)}$ for all possible realizations of fully separable states, bipartite entangled and tripartite entangled states, respectively, by following the procedures of Refs. \citen{Papp2009,Choi2010}.

Generally, we can determine the projection operators $\hat{\Pi}_i=|W_i\rangle\langle W_i |$ with $i\in\{1,\cdots,2^{m}\}$ for arbitrary number of systems $N_m=2^m$ with the recursive relationship,
\begin{equation}
|W_i^{(m)}\rangle=\frac{1}{\sqrt{2}} \left( |W_{i}^{(m-1)},\tilde{G}^{(m-1)}\rangle\pm | \tilde{G}^{(m-1)}, W_{i}^{(m-1)}\rangle \right),\nonumber
\end{equation}
from the initial condition $|W_{1,2}^{(1)}\rangle=1/\sqrt{2}(|gr\rangle\pm|rg\rangle)$. Here, $|W_i^{(m)}\rangle=(1/\sqrt{2^m})\sum_i^{2^m} |\tilde{r}_i^{(1)} \rangle$ and $|\tilde{G}^{(m)}\rangle=|g\cdots g\rangle$ for $N_m$ atoms. As discussed in Ref. \citen{Logouvski2009}, we then construct the uncertainty witness $\Delta=\sum_i \langle\delta^2 \hat{\Pi}_i\rangle$ to identify the bounds $\{\Delta^{(k-1)}_b\}$ for $(k-1)$-partite entanglement up to $k\leq N_m$. For convenience, we set the maximal $N_m\geq N_A$ to be larger than the number $N_A$  of atoms in $A$, so that we could distinguish the entanglement depth $k$ for any $k\leq N_A$.

For Fig. \ref{fig4}, we assumed the stationary limit, so that $y_c\rightarrow 0$. In order to verify the minimum bounds $\Delta_b^{(k-1)}$, we only need to optimize the overlap of pure $(k-1)$-partite entangled states of the form $|\psi_b^{(k-1)}\rangle=|\tilde{G}^{(N-k+1)}\rangle\otimes \sum_i^{k-1} \alpha_i |\tilde{r}_i^{(1)}\rangle$ with one of the projectors $|W_i\rangle$. This is achieved when the test state is a balanced $(k-1)$-partite $W$-state (i.e., $|\alpha_i|=1/\sqrt{k-1}$). Supplemtary Fig. 4 (a) depicts the uncertainty bounds $\Delta_b^{(k-1)}$ with $k\in\{1,\cdots,N_m\}$ calculated for $y_c=0$ and $N_m=2^7=128$. The shaded regions represent the parameter spaces for which ambiguity exists for the tiered structure $\Delta_b^{(k)}\not\geq\Delta_b^{(k+1)}$. This is caused by the nonlinear structure of $\Delta(|\psi_b^{(k)}\rangle\langle\psi_b^{(k)}|)$ to POVM values $\hat{\Pi}_i$. For such regions, we conservatively quote the minimum value of $k_m$ and certify the presence of genuine entanglement depth $k_m+1$ stored in the purported state $\hat{\rho}$ with $\Delta(\hat{\rho})<\Delta_b^{(k_m)}$ [See Supplementary Fig. 4(b)]. Hence, the entanglement depth $k_m$ (red dot) is a conservative estimate, which can be detected in an experiment, as opposed to the model-dependent analysis of $k$ (blue dot) for the pure state form $|\epsilon_1\rangle=|W_k\rangle_A\otimes |g\cdots g\rangle_B$ (i.e., by counting the number of non-zero probability amplitudes in $|W_k\rangle_A$).

Experimentally, the witness $\{\Delta(\hat{\rho}),y_c(\hat{\rho})\}$ can be determined by detecting the fluctuation $\delta^2 \vec{\mathcal{S}}_t$ in the collective transverse spin component $\vec{\mathcal{S}}_t=\sum_i \{\cos\theta_d\hat{\sigma}_x^{(i)},\sin\theta_d\hat{\sigma}_y^{(i)}\}$ and the excitation statistics $\{p_0,p_1, p_{\geq 2}\}$, where $\theta_d$ is the detection angle in the transverse plane $x-y$. As discussed in Ref. \citen{Choi2010}, 
\begin{eqnarray}
\Delta(\hat{\rho}) &\leq& \tilde{\Delta}(\hat{\rho}),\\
\tilde{\Delta}(\hat{\rho}) &=&\left (\frac{N-1}{N}\right )\times\left ( 1-N^2 \tilde{d}^2 \right),
\end{eqnarray}
where $\tilde{d}=\frac{2}{N(N-1)}\sum_{ij}|d_{ij}|$ is the average off-diagonal coherence $d_{ij}=|g\rangle_i\langle r| \otimes|r\rangle_j\langle g|$ for the reduced density matrix  $\hat{\rho}_1$ in the  single-excitation subspace. Since $\min \langle \delta^2 \mathcal{S}_t\rangle_{\theta_d}=2\sum_{ij}|d_{ij}|$, we find the following upper bound of the measured variance
\begin{equation}
\tilde{\Delta}(\hat{\rho})=\frac{N}{N-1}\times \left [ 1-\left (\frac{\min \langle \delta^2 \mathcal{S}_t\rangle_{\theta_d} }{N-1}\right)^2\right].
\end{equation}
The quantum statistics $y_c=\left (\frac{2N}{N-1}\right )\frac{p_{\geq 2} p_0}{p_1^2}$ can be detected by the total excitation statistics $\{p_0,p_1, p_{\geq 2}\}$ with MCP ionization signals. Hence, our entanglement witness can be readily implemented even for \textit{low-resolution} Rydberg experiments without the capability to locally detect the state of single atoms in the lattice.

\vspace{0.25cm}

\noindent\textbf{Experimental parameters with alkali atoms.} Let us consider $^{85}$Rb atoms interacting with optical field near the transition between $|g\rangle=|5S_{1/2}\rangle$ and $|r\rangle=|nS_{1/2}\rangle$ with two-photon Rabi frequency $\Omega={\Omega_1\Omega_2}/{\Delta^{\prime}}$ and detuning $\delta$ that globally addresses the atomic sample. As shown by Supplementary Fig. 5, this could be achieved by a two-photon transition with Rabi frequencies $\Omega_1,\Omega_2$ via the intermediate state $|e^{\prime}\rangle=|5P_{1/2}\rangle$ with one-photon detuning $\Delta^{\prime}$. The Rydberg excitation spectrum displays a highly nonlinear excitation spectrum $n$ due to the dipole-dipole interaction $\Delta_p^{(ij)}=C_p|\vec{r}_i-\vec{r}_j|^{-p}$, with the most shifted level given by configuration states consisting of nearest-neighbor excitations $|\tilde{r}_{i,i+1}^{(2)}\rangle$ with $\Delta_p^{(i,i+1)}$. 

In order to achieve the parameter sets of Figs. \ref{fig2}--\ref{fig4}, we take the principal quantum number $n_p=100$ so that $|r\rangle=|100S_{1/2},m_j=1/2\rangle$. The radiative lifetime is given by $\tau=\tau_0 {(n_p^{*})}^\alpha$, where $n_p^{*}=n_p-\delta_{nl}$ is the effective principal number and $\delta_{nl}$ is the quantum defect. With $\tau_0=1.43$ ns and $\alpha=2.94$ for $|100S_{1/2}\rangle$ (Ref. \citen{Rydbergbook}), we find that the Rydberg lifetime is $\tau_r=1$ ms (i.e., $\Gamma_r\simeq 1$ kHz). On the other hand, $\Gamma_e\simeq 36 $ MHz for $|e\rangle$. Since $\Gamma\rightarrow \Gamma_e$ in the limit of strong dressing fields $\Omega_1$ for ``reservoir" atoms, $\Gamma\rightarrow 10^4\Gamma_r$ can be achieved in an experiment. 

By setting $\Omega_{1,2}=100$ GHz and $\Delta^{\prime}=10\Omega_{1,2}$ (i.e., $\Omega=10$ GHz), the photo-ionization rate can be determined by
\begin{equation}
\gamma_{\pi}=\frac{I}{\hbar w }\times \sigma_{\pi}=\frac{2I_{\text{sat}}}{\hbar w}\left( \frac{\Omega_{1,2}}{\Gamma_e^{\prime}} \right)^2 \sigma_{\pi},
\end{equation}
where $\Gamma_e^{\prime}= 36$ MHz is the spontaneous decay rate for $|e^{\prime}\rangle$, $I_{\text{sat}}=4.5$ mW/{cm}$^2$ is the saturation intensity for $|g\rangle \rightarrow |e'\rangle$ and $\sigma_{\pi}\leq 2\times 10^{-7} \AA^2$  is the photo-ionization cross-section that couples the Rydberg state $| 100S_{1/2}\rangle$ to the continuum free-electron wavefunctions \cite{Rydbergbook}. Hence, we find that the photo-ionization lifetime is limited to $\tau_{\pi}=1/\gamma_{\pi}\gg 4$  ms $\gg 1/\Gamma_r$ (Ref. \citen{Saffman2005}).

The blockade shift $\Delta_{p}^{(ij)}$ is determined by Rydberg coefficient $C_p$, for which we take $C_6=56$ THz$\cdot \mu \text{m}^6$ for the vdW interaction between two Rydberg atoms in $|r\rangle=|100S_{1/2},m_j=1/2\rangle$ (Refs. \citen{Singer2005,Dudin2010,Balewski2013}). For $\Omega=10$ GHz and $N=4,6$, the blockade shift for nearest-neighbors is $\Delta^{(i,i+1)}_{6}=20$ THz ($a_0/d_B\simeq 0.3$), while the power-broadened linewidth for the transition $|g\rangle\leftrightarrow |r\rangle$ is $w_d\simeq \sqrt{2}\Omega$. The resulting blockade radius is $r_B\simeq \sqrt[6]{C_6/w_d}=4 \mu$m and $1(1.2) \mu\text{m}\leq a_0(a_1)\leq 1.5 (1.8)\mu\text{m}$ for $F_4> 0.99$. Even for $N=128$ atoms, in which $a_0/d_B\simeq 0.13$, we can set $\Omega=10^2\Gamma$, $\Gamma/\Gamma_r=10^4$ and $\xi=a_1/a_0=\sqrt[6]{3}$, thereby leading to the blockade distance $d_B=\sqrt[6]{C_p/w_d}=5.8\mu $m and $(a_0,a_1)\simeq (750\text{ nm},900\text{ nm})$. In terms of spatial localizations, the variance of the lattice constants would need to be less than $\delta a_0, \delta a_1< 150$ nm in order to achieve $F>0.99$ in Figs. \ref{fig2}--\ref{fig4}. This could be readily achieved in deep optical lattice experiments with zero-point motion $\delta x\sim 10$nm. Hence, Rydberg atoms interacting in the strong blockade regime with the lattice constants $a_0,a_1\simeq  1 \mu$m $>\lambda_0/2$ (Figs. \ref{fig2}--\ref{fig4}) can be spatially resolved, so that $\Omega_d$ can be locally addressed to the reservoir sites without the requirement for diffraction-limited imaging resolutions $\lambda_0/2$ (Ref. \citen{Bakr2009}).

Therefore, the pumping time for $F>0.95$ for Figs. \ref{fig2}--\ref{fig3} is then $\tau_p\sim 10^2/\Gamma=10\mu \text{s}$, which is not limited by the photo-ionization time $\tau_{\pi}\gg 4$ ms. In addition, since the quantum jumps in the $n=1$ subspace occur on a time-scale of $t_j\sim \mathcal{O}(N^2)$ due to the random walk for $|\tilde{r}_i^{(1)}\rangle$ until it reaches the ``reservoir" sites with $\Gamma_{1,N}=\Gamma\gg \Gamma_r$, we expect that the pumping time to reach stationarity also scales as $t_p\sim  \mathcal{O}(N^2)$. On the other hand, if we were to address every ``zeros" in Eqs. \ref{set1eqs}--\ref{set2eqs} with $\Omega_d$, the pumping time $t_p\sim  \mathcal{O}(N)$ will scale linear to the number $N$ of eigenstates $\{|\epsilon_{\mu}\rangle\}$ spanning $n=1$. By extrapolation, we estimate that our method could be extended to generate $100$-partite entangled steady-states with the parameters $\{\Omega_{1,2},\delta,\Delta^{\prime},|g\rangle,|e^{\prime}\rangle,|r\rangle\}$. Further improvement in the entanglement depth $k$ may be possible by optimizing the driving field $\Omega$ under the constraint $\frac{2\Omega^2}{\delta}\gg \Gamma_r$ for a given $|r\rangle$, which reduces the ionization time $t_{\pi}$ (Ref. \citen{Saffman2005}). Alternative strategies, including the use of photonic crystals with atoms in low-lying electronic states, will be discussed elsewhere.

In terms of the initialization of the atoms in the 1D lattice, the atoms would need to be confined in each well with unit filling factor. In practice, this could be achieved by the superfluid-Mott insulator transition or by the manipulation of laser-induced atomic collisions with blue-detuned potentials \cite{Grunzweig2010}. The 1D staggered triangular lattice can be easily realized in a free-space superlattice configuration \cite{Jo2012}. Since the general principle of our protocol is not necessarily confined to a particular lattice configuration, one could explore other configurations in 1D and 2D with arbitrary trap potential landscapes created by spatial light modulators (SLM).}

\clearpage

\title{Supplemental Figures for  \\``Emergence of stationary many-body entanglement in driven-dissipative Rydberg lattice gases"}
\author{Sun Kyung Lee}
\address{Spin Convergence Research Center, Korea Institute of Science and Technology, Seoul 136-791, Korea}
\author{Jaeyoon Cho}
\address{School of Computational Science, Korea Institute for Advanced Study, Seoul 130-722, Korea}
\author{K. S. Choi}
\address{Spin Convergence Research Center, Korea Institute of Science and Technology, Seoul 136-791, Korea}
\address{Institute for Quantum Computing and Department of Physics \& Astronomy, University of Waterloo, Waterloo, Ontario N2L 3G1, Canada}
\maketitle


\begin{figure*}[h]
\includegraphics[width=1\columnwidth]{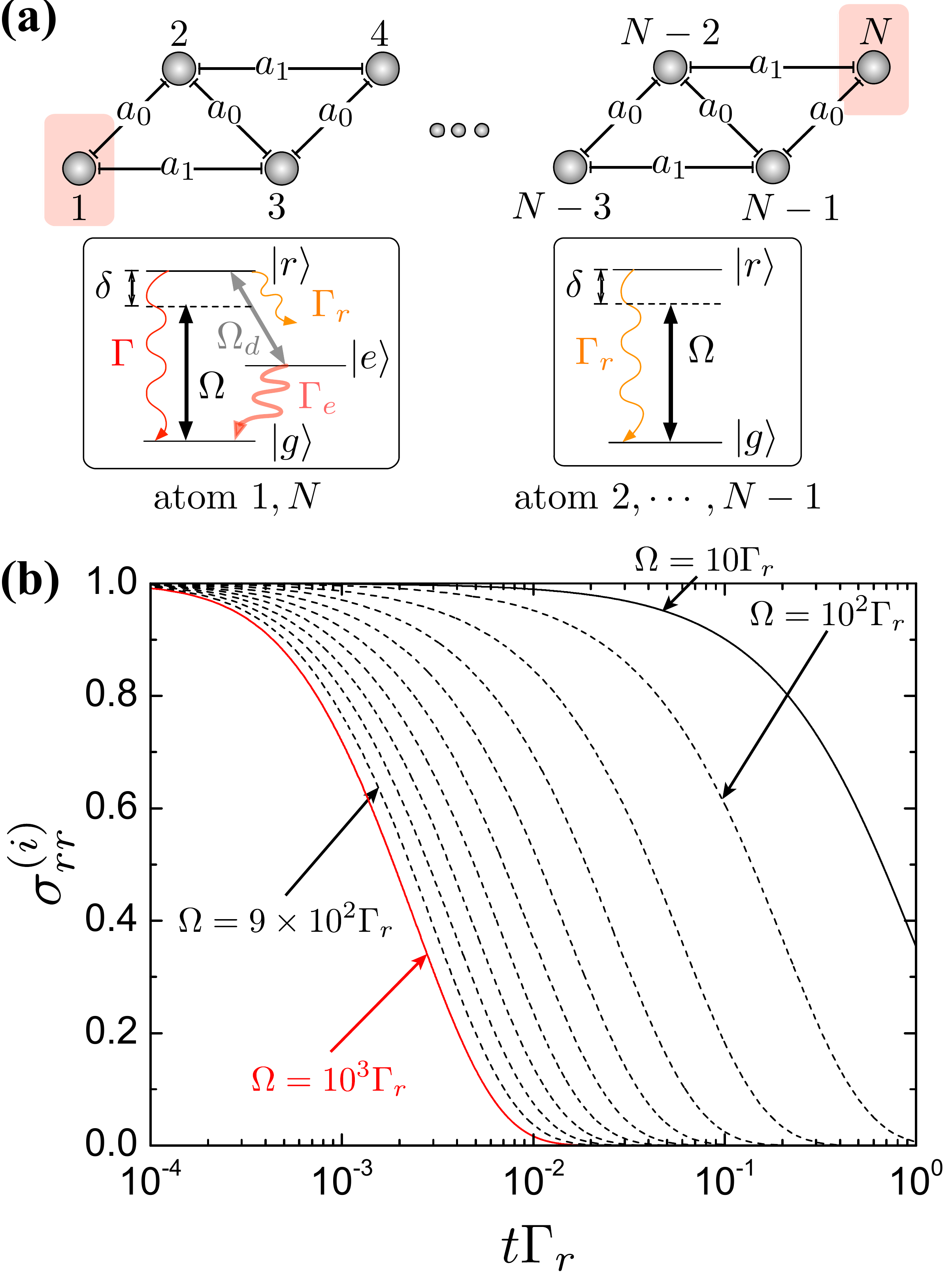}\\ 
\caption{\textbf{(a)} Control of spontaneous emission rate via dressing field. Inset
\textbf{(i)} The decay rates for the edge atoms are enhanced by dressing the Rydberg states $|r\rangle$ with auxiliary excited state $|e\rangle$ via the dressing fields $\Omega_d$. Inset \textbf{(ii)} Atoms are pumped by the driving field $\Omega$ with detuning $\delta$. \textbf{(b)} Enhancement in the decay rate $\gamma_{\text{eff}}\simeq |\Omega_d|^2/\Gamma_e$ as a function of the strength of dressing field $\Omega_d$. The parameters are $\Gamma_e=10^4\Gamma_r$, and resonant dressing $\Delta_d=0$.} \label{enhanced} 
\end{figure*}

\newpage
\begin{figure*}[h]
\begin{center}
\includegraphics[width=1\columnwidth]{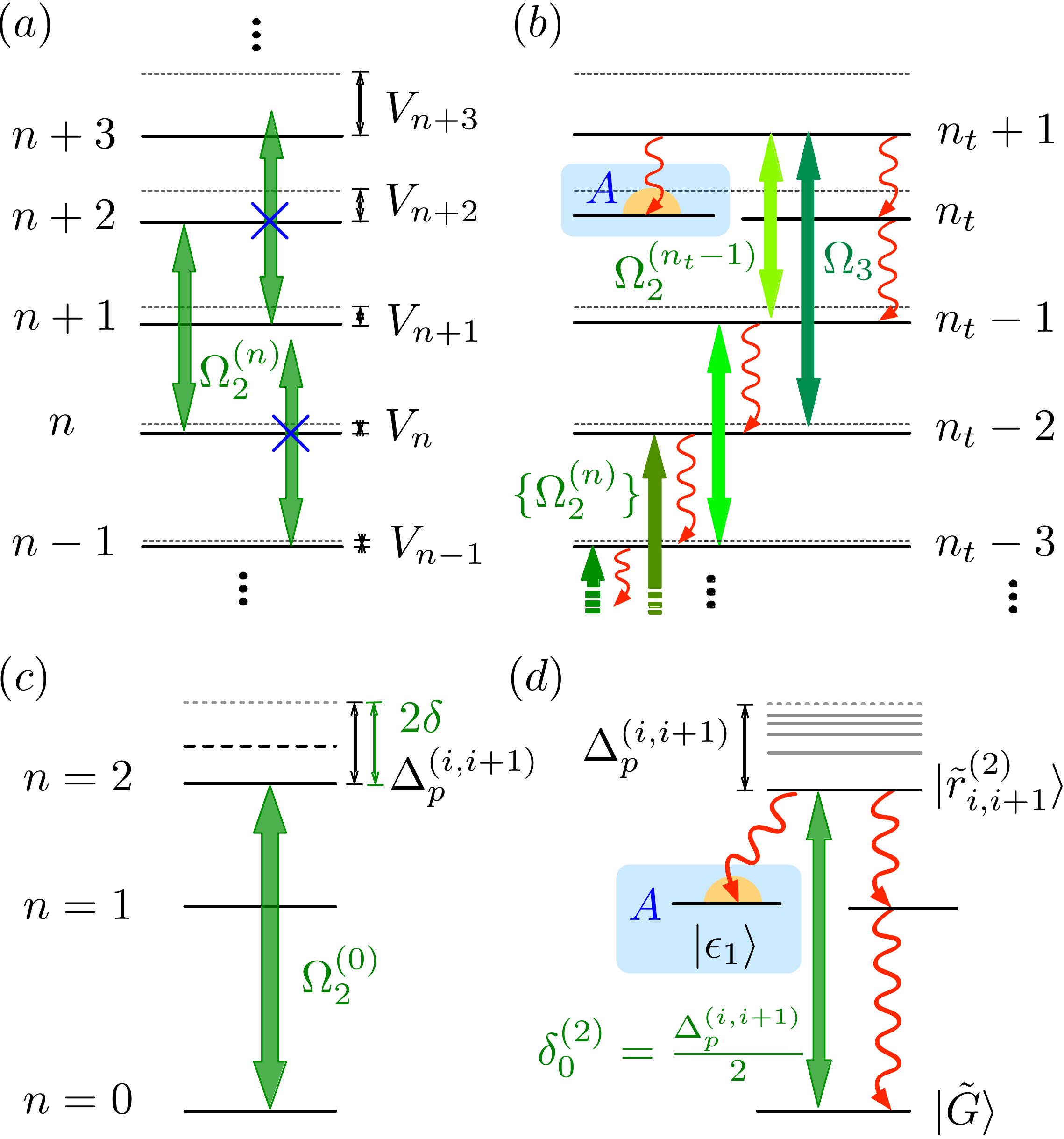}
\caption{ \textbf{(a)} Anharmonic Rydberg spectrum for two-photon transition $\Omega_2^{(n)}$. Dotted (solid) line represents the energy level for non-interacting atoms (the most shifted energy level with Rydberg interaction). \textbf{(b)} A set of $n_t-1$ two-photon transitions $n\rightarrow n+2$ (with $n \in \{0,\cdots, n_t-1\}$) are resonantly driven at Rabi frequencies  $\{\Omega_2^{(n)}\}$, in tandem with a three-photon laser $\Omega_3$ for $n_t-2\rightarrow n_t+1$ are required to pump atoms to the target subspace $n_t$. Non-coupled subspace $A$ of $n$ excitations remains dark throughout the entire driving processes. The target eigenstate is $|\epsilon_1\rangle=\sum_{\{p_i\}\in A} \alpha_{\{p_i\}} |\tilde r^{(n)}_{\{p_i\}}\rangle_A \otimes |g\cdots g\rangle_B$. \textbf{(c)} A single laser is required to drive $n=0\rightarrow n=2$ for target subspace $n_t=1$,  \textbf{(d)} For $\delta=\delta_0^{(2)}$, atoms are pumped to the target eigenstate $|\epsilon_{1}\rangle=\sum_{i\in A} \alpha_{i,1} |\tilde r^{(1)}_{i}\rangle_A\otimes |g\cdots g\rangle_{B}$, whereby decoherence is enhanced for atoms $B$.}\label{arbnpumping}
\end{center}
\end{figure*}

\newpage

\begin{figure*}[h]
\begin{center}
\includegraphics[width=1\columnwidth]{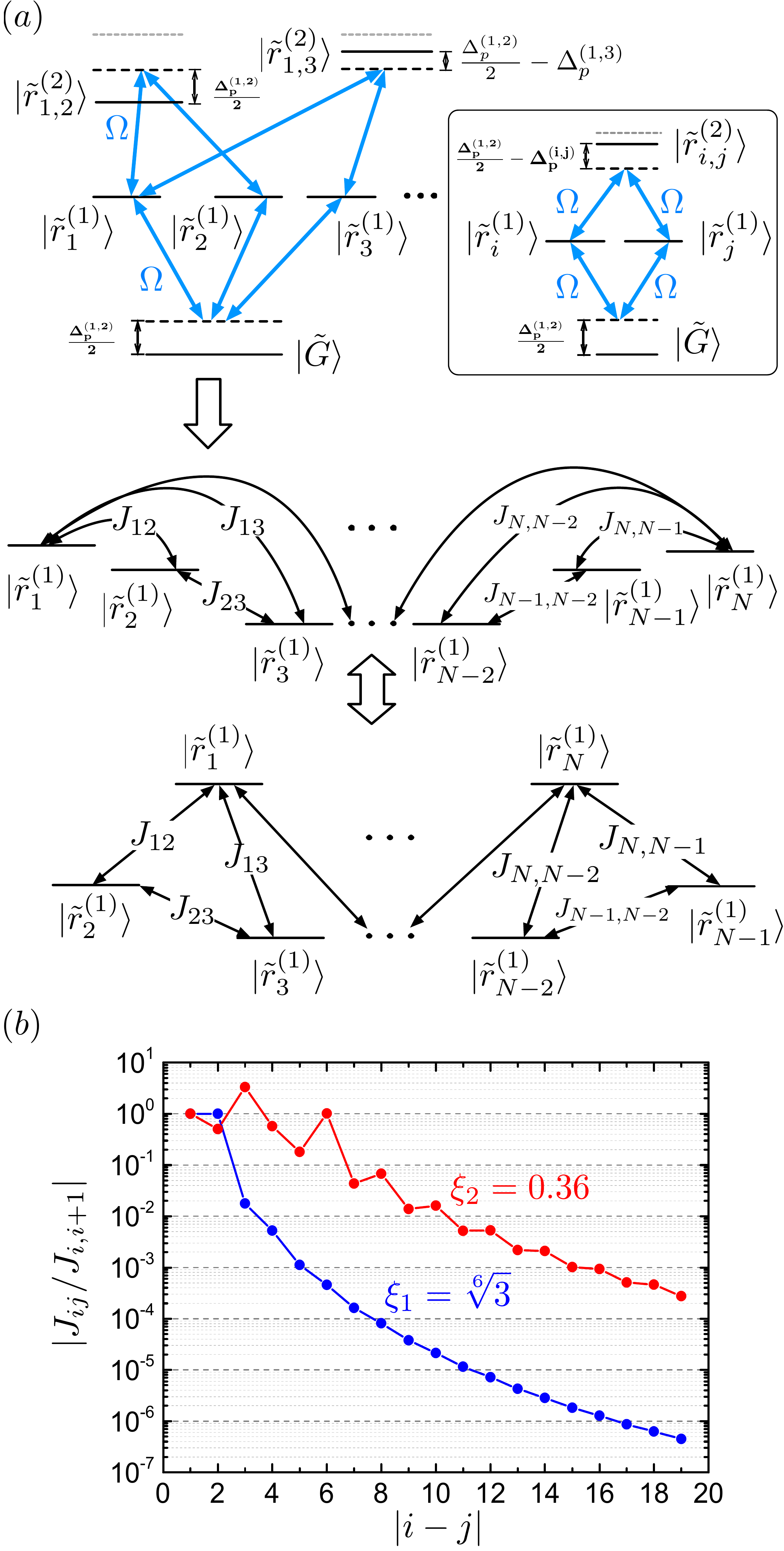}
\caption{\textbf{(a)} Interaction strength $J_{ij}$ and light shift $\bar\Delta_{\text{ls}}^{(i)}$ of the effective Hamiltonian $H_{xy}$ are given by Raman transition $|\tilde r_{i}^{(1)}\rangle \leftrightarrow |\tilde r_{j}^{(1)}\rangle$ via two path mediated by $|\tilde G \rangle$ and $|\tilde r_{i,j}^{(2)}\rangle$. \textbf{(b)} The power law scaling behavior of spin-spin coupling strength $J_{ij}$ with 1D staggered triangular lattices for $\xi_1=\sqrt[6]{3}$ (blue) and $\xi_2\simeq 0.36$ (red). For $\xi_1$, the spatial range of $J_{ij}$ depicts a monotonic power-law decay, whereas jig-jag oscillatory pattern exists for $\xi_2$.}\label{figJij}
\end{center}
\end{figure*}

\newpage

\begin{figure*}[h]
\includegraphics[width=1\columnwidth]{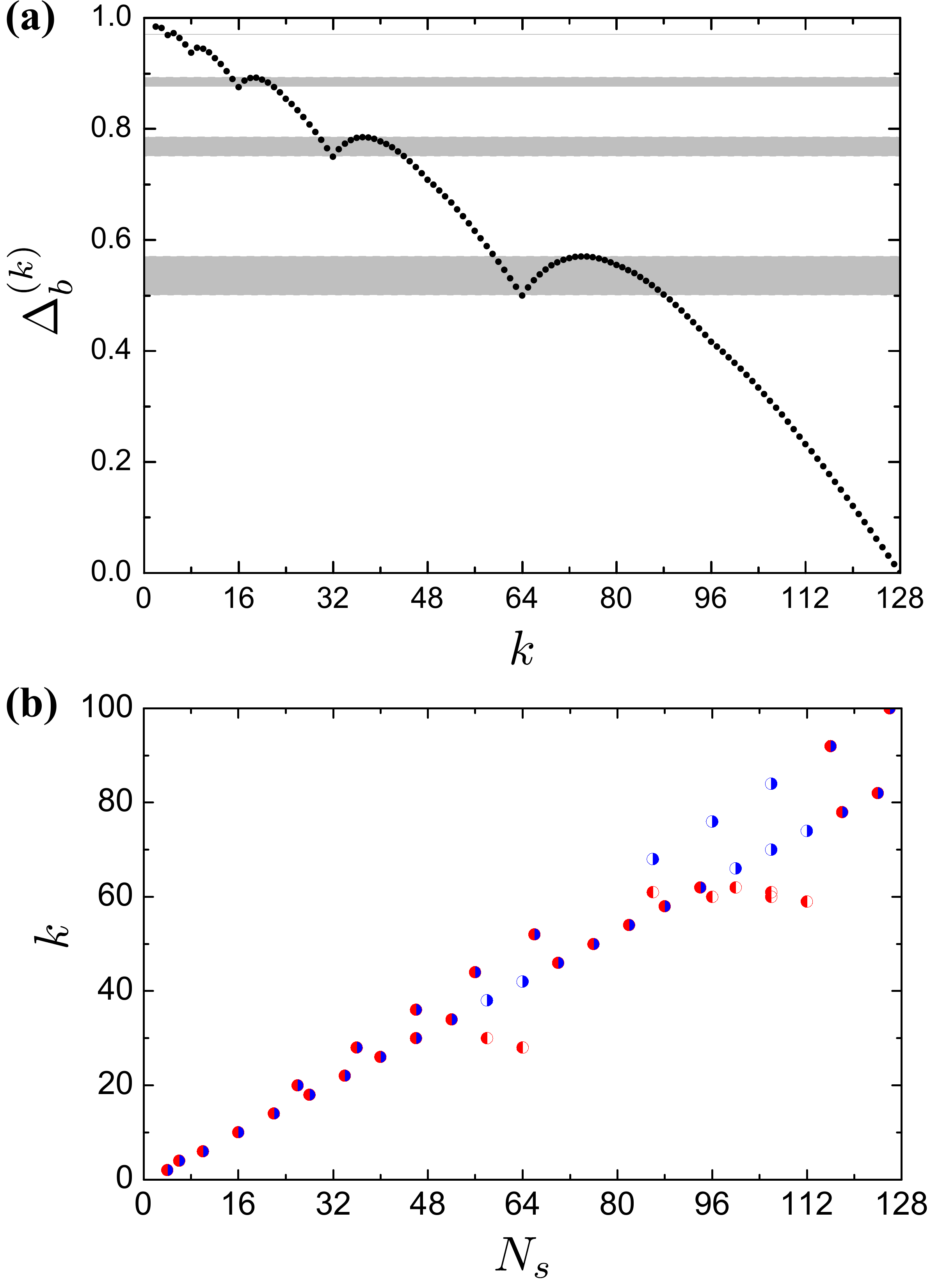}\\ 
\caption{\textbf{(a)} $k$-partite uncertainty bounds $\Delta_b^{(k)}$ for $k\in\{1,\cdots,128\}$ and $N_m=128$. Shaded regions indicate the parameter spaces for which ambiguity exists for $\Delta_b^{(k)}\not\geq\Delta_b^{(k+1)}$, due to the nonlinear sensitivity of $\Delta$. For such regions, we conservatively quote the minimum value $k_m$ for the genuine $k$-partite entanglement with $\Delta(\hat{\rho})<\Delta_b^{(k_m)}$. \textbf{(b)} The minimum entanglement depth $k_m$ certified by $\{\Delta,y_c\}$ (red dots), and the entanglement depth $k$ in the purported eigenstate $|\epsilon_1\rangle=|W_k\rangle_A\otimes |g\cdots g\rangle_B$ (blue dots), with fully balanced $k$-partite $W$ state $|W_k\rangle_A$. } \label{bounds} 
\end{figure*}

\newpage

\begin{figure*}[h]
\includegraphics[width=1\columnwidth]{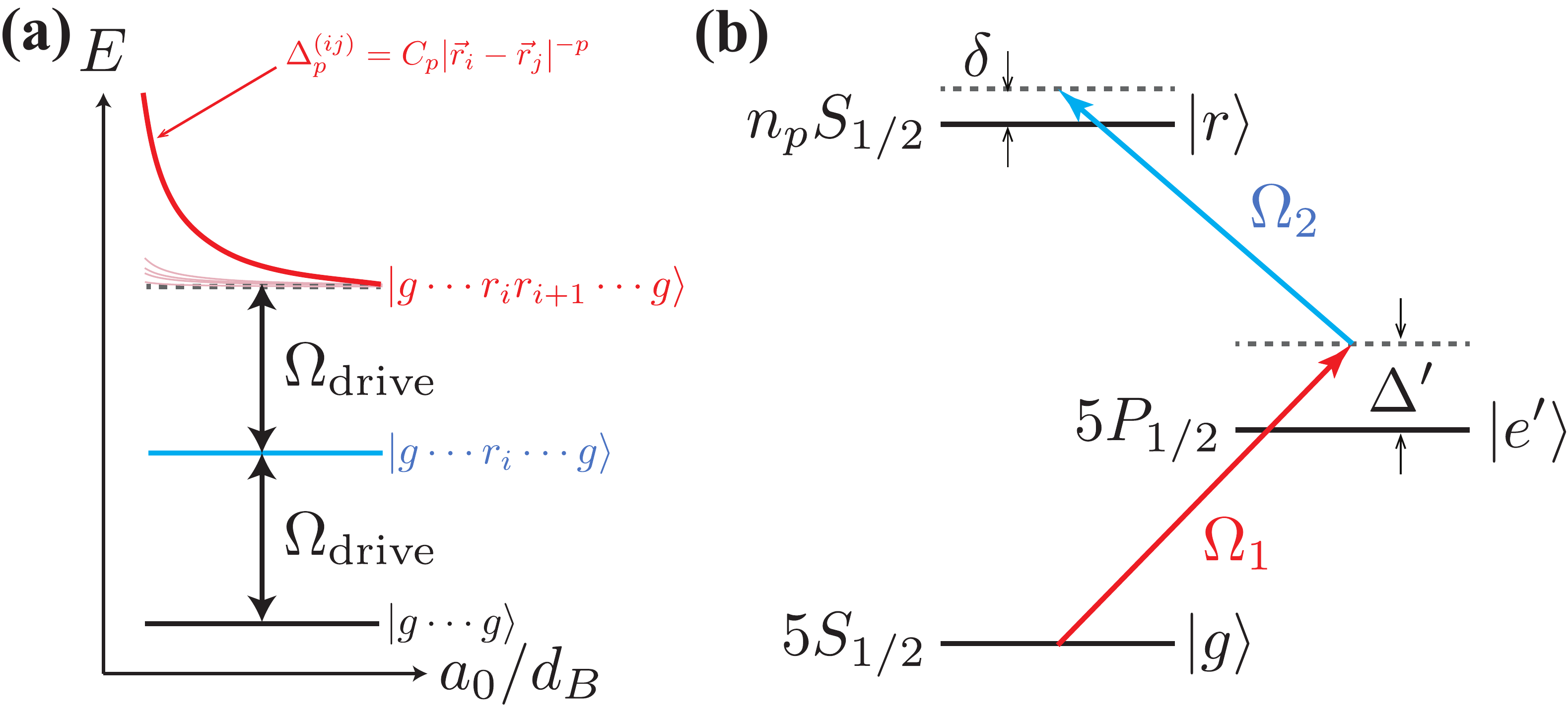}\\ 
\caption{\textbf{(a)} $N$-atom Rydberg blockade. Effective Rabi frequency between $|g\rangle$ and $|r\rangle$ is given by $\Omega$. \textbf{(b)} Level diagram for $^{85}$Rb atom. The effective transition between $|g\rangle$ and $|r\rangle$ is formed by a two-photon transition via the intermediate excited state $|e^{\prime}\rangle$, with $|g\rangle=|5S_{1/2}\rangle$, $|e^{\prime}\rangle=| 5P_{1/2}\rangle$, and $|r\rangle=| n S_{1/2}\rangle$. $\Delta^{\prime}$ is the one-photon detuning respect to $|e^{\prime}\rangle$ by field $\Omega_1$ ($\lambda_1\simeq 474$ nm) and $\delta$ is the two-photon detuning by the field $\Omega_2$ ($\lambda_2\simeq 795$ nm).} \label{Rydberg-level-diagram} 
\end{figure*}


\end{document}